\documentclass[10pt,a4paper]{article}
\usepackage{amsmath}
\usepackage{amsfonts}
\usepackage{epsfig}
\usepackage{times}
\newcommand{\be}{\begin{equation}}
\newcommand{\ee}{\end{equation}}
\newcommand{\ba}{\begin{array}}
\newcommand{\ea}{\end{array}}
\newcommand{\bea}{\begin{eqnarray}}
\newcommand{\eea}{\end{eqnarray}}

\newcommand{\rar}{\rightarrow}
\newcommand{\p}{\partial}
\newcommand{\ol}{\overline}
\newcommand{\ti}{\tilde}
\newcommand{\la}{\langle}
\newcommand{\ra}{\rangle}
\newcommand{\bct}{\begin{center}}
\newcommand{\ect}{\end{center}}

\renewcommand{\l}{\newline\null}
\def\figskip{\vskip .4cm plus 2mm minus 2mm}
\def\hbar{h\!\!\!/}
\begin{document}
\begin{titlepage}
October 1998 (revised April 1999)\hfill PAR-LPTHE 98/21
\vskip 4.5truecm
{\baselineskip 17pt
\begin{center}
{\bf LARGE BREAKING OF ISOSPIN SYMMETRY IN AN ELECTROWEAK
$\mathbf{SU(2)_L \times U(1)}$
GAUGE THEORY OF $\mathbf {J =0}$ MESONS: THE CASE OF
$\mathbf {K \rar \pi\pi}$ DECAYS}
\end{center}
}
\vskip .5cm
\centerline{B. Machet
     \footnote[1]{Member of `Centre National de la Recherche Scientifique'.}
     \footnote[2]{E-mail: machet@lpthe.jussieu.fr.}
     }
\vskip 5mm
\centerline{{\em Laboratoire de Physique Th\'eorique et Hautes Energies,}
     \footnote[3]{LPTHE tour 16\,/\,1$^{er}\!$ \'etage,
          Universit\'e P. et M. Curie, BP 126, 4 place Jussieu,
          F 75252 PARIS CEDEX 05 (France).}
}
\centerline{\em Universit\'es Pierre et Marie Curie (Paris 6) et Denis
Diderot (Paris 7);} \centerline{\em Unit\'e associ\'ee au CNRS UMR 7589.}
\vskip 1.5cm
{\bf Abstract:} $K\rar \pi\pi$ decays are investigated in the framework of
the $SU(2)_L \times U(1)$ gauge theory of $J=0$ mesons proposed in
\cite{Machet1}, to which is added an interaction between mesons respecting the
two symmetries generally attributed to strong interactions,
flavour symmetry and parity conservation.
It is shown that the damping of the $K^+\rar \pi^+\pi^0$ amplitude
with respect to the ones of $K_s\rar \pi\pi$  results from a cancelation
between the  $W$ and $Z$ gauge bosons.
The ratio of the two types of amplitudes is sensitive to the spectrum of
electroweak (pseudoscalar and scalar) mass eigenstates, and in particular
to the mass of the Higgs boson. In the case where scalars and pseudoscalars
are degenerate, I demonstrate for the amplitudes  the lower bound
$\vert K_s \rar \pi^+\pi^-/K^+ \rar \pi^+\pi^0 \vert \geq 1/\tan^2\theta_W$,
where $\theta_W$ is the Weinberg angle.
\smallskip

{\bf PACS:} 12.15.Ji\ 12.15.Lk\ 12.60.-i\ 12.90.+b\ 13.25.Es\ 14.80.Bn\ 14.80.Cp
\vskip 1.5 truecm
\vfill
\null\hfil\epsffile{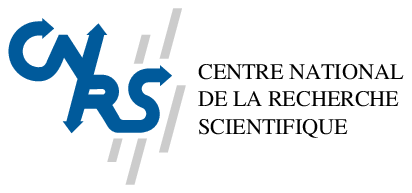}
\end{titlepage}
%
%
\section{Introduction}
\label{section:intro}

Large departures from isospin symmetry are observed in nature. They have made
in particular the puzzle of hadronic $K \rar \pi\pi$ decays \cite{RPP}
one among the oldest in particle physics, and the huge literature
devoted to the so-called  $\Delta I = 1/2$ rule covers 30 years of
investigations in this field \cite{Cheng}
\footnote{I shall not quote here all papers that have been devoted to the
subject, and shall rather mention works characterizing the main steps in the
evolutions of the ideas; the reader is conveyed to the extensive review
\cite{Cheng} for additional references.}
.

If electromagnetic and weak interactions, treated perturbatively,
can correctly account for a large variety of physical processes,
they are, alone, at a loss for $K \rar \pi\pi$ decays and
the natural idea was to incorporate, in one way or another, strong 
interactions.

The first attempts \cite{GellMannetCA} made use of Current Algebra
\cite{CurrentAlgebra}, a non-perturbative
technique based on exact commutation relations between the generators of
chiral $U(N)_L \times U(N)_R$, linked to currents the divergence of which
can be used as interpolating fields for the mesons; however, its performance is
maimed by the necessity of working at the ``soft'' limit,
far from the physical kinematic domain of the decays.

Then came out the renormalizable  unbroken colour gauge theory of quarks
and gluons \cite{QCD}; the techniques of operator product expansion
showed that gluonic exchanges enhanced the
coefficients of the ``octet'' four-quarks operators corresponding to $\Delta I
=1/2$ transitions \cite{GaillardLee}; the discovery of the so-called
``penguin diagrams'' \cite{SVZ} provided a new perspective on the subject.
Still, though most  computations go in the right direction,
the enhancement is in
general  too small and difficulties cannot be concealed that:\l
- the large gap already mentioned in \cite{Feynman} between Quantum
Chromodynamics and a theory for the strong interactions of hadrons still
exists, and
the Yang-Mills theory of colour has to be supplemented with various hypotheses
concerning confinement and hadronization (PCAC, soft limit, factorization
and vacuum insertion \cite{BonvinSchmid} {\em etc});\l
- at the scales involved in $K \rar \pi\pi$ decays, it is presumably not
a weakly coupled theory and treating it perturbatively or in the
one-gluon approximation is problematic.\l
The intractable complexity of higher order computations can be slightly
decreased by reordering the expansion in inverse powers of the number of
colours \cite{1/N};
this technique culminated in the works \cite{BardeenBurasGerard}.
If, again, the results go in the right direction, no rapid improvement
can be expected because of the intricacies of the calculations.

The techniques of effective theories \cite{chiralth} incorporating chiral
properties of mesons and some of the features of Quantum Chromodynamics have
also been applied to this problem \cite{Kpipichiral}; they can in particular
prepare the path to computer simulations.

The sole truly non-perturbative technique available is indeed putting the system
on a lattice and making computer simulations.  It has been applied to $K\rar
\pi\pi$ decays, and more specially to $\la \pi\vert K\ra$ matrix elements
after using Current Algebra to ``reduce'' one of the two pions \cite{lattice}.
It is not either devoid of ambiguities.

If many of these techniques provide encouraging results,
none seems to be really convincing and physically simple enough to be
appealing; the great complexity of the computations and the
various hypotheses that must be been introduced often dim the demonstrations.

A common belief is that the $\Delta I = 1/2$ rules for kaon decays results
from the combined action of many different mechanisms, and that a unique and
transparent interpretation may always be lacking.

The standard electroweak model \cite{GlashowSalamWeinberg} is not by itself
devoid of unanswered and
fundamental questions; in particular, the renormalizable generation of mass
is at present only performed through the mechanism of spontaneous symmetry
breaking, making use of the most elusive Higgs boson.
Decoupling theorems \cite{decoupling} show that a fundamental Higgs
is perturbatively highly protected.
If finding it would be a great achievement, it is not even
sure whether it should indeed be considered as a fundamental particle, and
other directions of investigations are worth considering \cite{Higgs}.

In the traditional framework of the Glashow-Salam-Weinberg model
\cite{GlashowSalamWeinberg}, neither the Higgs nor its three partners in
the complex scalar doublet, which are the goldstones of the broken
electroweak symmetry, bear any connection with the observed mesons;
in particular, in the strict framework of the standard model, the former
plays no sensitive role in mesonic decays. 

I investigate here $K \rar \pi \pi$ decays in the framework of the
renormalizable
electroweak $SU(2)_L \times U(1)$ gauge theory of $J=0$ mesons proposed in
\cite{Machet1}; they are considered to be both the fundamental fields of the
Lagrangian and the asymptotic states, and to transform like $(\bar q_i q_j)$
or $(\bar q_i \gamma_5 q_j)$ operators by the chiral group $U(N)_L
\times U(N)_R$ ($N$ is the number of flavours) and by its electroweak
subgroup; the orientation of
the latter into the former is unambiguously determined by the identification
of the $q_i$'s with the $N$ quarks on which also acts the gauge group of the
electroweak standard model. Both behaviors of the quarks,
by chiral and electroweak transformations, are thus incorporated.

In addition to the status given to quarks, one of the main differences
with the standard approach lies in that the
Higgs boson and its three pseudoscalar (goldstone) partners now form one
among the $N^2/2$
quadruplets into which the $2N^2$ electroweak $J=0$, scalar and pseudoscalar,
mesonic eigenstates can be classified. They are all linear combinations of
flavour or ``strong'' eigenstates ($\pi, K, D, D_s \ldots$ and their scalar
equivalent).

As a consequence, electroweak interactions now couple the Higgs boson to
those of the pseudoscalar mesons which build up the three goldstones:
this distinguishes the charged from the neutral kaons and introduces from
the start an asymmetry between the two types of incoming states occurring in
$K \rar \pi\pi$ decays.
The Higgs boson appears in internal lines of 1-loop electroweak diagrams
controlling $K^+ \rar \pi^+ \pi^0$ decays and does not in the decays of
neutral kaons.
This effect is however negligeable, as shown in the first part of this work.

It turns out, too, that no $\pi^0\pi^0$ final state can occur in the pure
electroweak decays of $K_s$ mesons. This drastic breaking of strong isospin
symmetry, of pure electroweak origin, adds up to the problem of accounting
for the damping of $K^+ \rar \pi^+\pi^0$ decays and makes
mandatory the introduction of another type of interactions.

This is performed in the second past of the work where I propose a very
simple model of interactions between mesons which  respect
flavour symmetry and conserves parity. The (unknown) coupling constant
$\Lambda$ drops out from the ratio of amplitudes, making our estimates for
the latter independent of the strength of these interactions.

The combined action of electroweak interactions and $\Lambda$-interactions
for mesons restores, at the one-loop order, the occurrence of
$K_s \rar \pi^0\pi^0$ decays, and provides an elegant and simple mechanism for
the damping of $K^+\rar \pi^+\pi^0$ amplitude
with respect to $K_s \rar \pi\pi$: while the latter involve only
contributions from the $W$ gauge boson, both $W$ and $Z$ occur in the former
and their contributions tend to cancel.
\footnote{$W-Z$ cancelation had already been shown
\cite{GlashowSchnitzerWeinberg} to provide finiteness of
the current algebra computation of $K \rar \pi\pi$ decays with the help of
spectral function sum rules.}
There is no obstacle to reproduce the observed ratios of amplitudes, and the
cancelation just mentioned can even be complete.
The computation of the amplitudes depends on the 
spectrum of electroweak mass eigenstates which propagate in internal lines,
including the scalar mesons; as it is for a large part unknown, we have to
work within certain approximations:
the crudest takes all electroweak eigenstates to be degenerate;
the intermediate one splits scalars and pseudoscalars;
the finest (though still crude) introduces two more mass scales,
the vanishing mass of the three (pseudoscalar) goldstones and an independent
mass for the Higgs boson, which makes up finally four different mass scales.
This last approximation exhibits the sensitivity of the $W-Z$ cancelation to
the electroweak spectrum, in particular to the mass of the Higgs boson. This
means that, would this be the only unknown, the experimental value of the
ratio of $K^+ \rar \pi^+\pi^0$ and $K_s\rar\pi^+\pi^-$ amplitudes would
yield an estimate for it.

More refined approximations are unrealistic in our present state of
knowledge of the spectrum of electroweak eigenstates (specially scalars),
and are therefore out of the scope of the paper.
  
\section{Theoretical framework: electroweak interactions of quark-antiquark
composite fields}
\label{section:theory}

The general framework has been set in \cite{Machet1}. For the sake of
understandability and for this paper to be self-contained, I briefly
recall here the main useful steps, in a somewhat less formal approach more
usable for phenomenological purposes.

Quarks are considered to be mathematical objects \cite{GellMann}
which are determined by their quantum numbers and by their transformations
by the different groups of symmetry that act upon them; we are mainly
concerned here with the chiral group $U(N)_L \times U(N)_R$ where $N$ is
the number of ``flavours'', and with the electroweak group
$SU(2)_L \times U(1)$: they form an $N$-vector $\Psi$
\begin{equation}
\Psi =
\left(
\ba{c}  u\\ c\\ \vdots \\d\\ s\\ \vdots \ea
\right)
\label{eq:Psi}\end{equation}
in the fundamental representation of the diagonal subgroup of the chiral
group,
and their electroweak transformations, to which we stick to, are the usual
ones of the Glashow-Salam-Weinberg model \cite{GlashowSalamWeinberg}.

Any $SU(2)_L \times U(1)$ group can be considered, for $N$ even, as a
subgroup
of $U(N)_L \times U(N)_R$; that it be the electroweak group, {\it i.e.}
that it act on quarks in the standard way determines its embedding
in the chiral group.
It is easy to check that the $SU(2)$ group the three generators of which are
the three $N \times N$ matrices (they are written in terms of four $N/2
\times
N/2$ sub-blocs with $\mathbb I$ the unit matrix and $\mathbb K$
the Cabibbo-Kobayashi-Maskawa (CKM) mixing matrix \cite{CKM})
\begin{equation}
{\mathbb T}^3_L = {1\over 2}\left(\begin{array}{rrr}
                        {\mathbb I} & \vline & 0\\
                        \hline
                        0 & \vline & -{\mathbb I}
\end{array}\right),\
{\mathbb T}^+_L =           \left(\begin{array}{ccc}
                        0 & \vline & {\mathbb K}\\
                        \hline
                        0 & \vline & 0           \end{array}\right),\
{\mathbb T}^-_L =           \left(\begin{array}{ccc}
                        0 & \vline & 0\\
                        \hline
                        {\mathbb K}^\dagger & \vline & 0
\end{array}\right),
\label{eq:SU2L}
\end{equation}
acting trivially on the left-handed projection $\Psi_L = [(1-\gamma_5)/2]
\Psi$
of $\Psi$ can be identified with the standard electroweak $SU(2)_L$;
the $U(1)$ associated to the weak hypercharge $\mathbb Y$  is determined
through the Gell-Mann-Nishijima relation \cite{GellMannNishijima}
\begin{equation}
({\mathbb Y}_L,{\mathbb Y}_R) =
                       ({\mathbb Q}_L,{\mathbb Q}_R) - ({\mathbb T}^3_L,0),
\label{eq:GMN}\end{equation}
and from the trivial form for the (diagonal) charge operator $\mathbb Q$
\begin{equation}
{\mathbb Q}_L ={\mathbb Q}_R ={\mathbb Q}=\left(\begin{array}{ccc}
                        2/3 & \vline & 0\cr
                        \hline
                        0 & \vline & -1/3
           \end{array}\right).
\label{eq:Q}
\end{equation}
The $2N^2$ composite of the form $\bar q q$ or $\bar q\gamma_5 q$ can be
cast
into $N^2/2$ quadruplets which are stable by the electroweak group; their
flavour structure is materialized by $N\times N$ matrices
$\mathbb M$ and the quadruplets can generically be written

\vbox{
\bea
& &\Phi(\mathbb D)=
({\mathbb M}\,^0, {\mathbb M}^3, {\mathbb M}^+, {\mathbb M}^-)(\mathbb D)\cr
& &\ \cr
& & =\left[
 \frac{1}{\sqrt{2}}\left(\begin{array}{ccc}
                     {\mathbb D} & \vline & 0\\
                     \hline
                     0 & \vline & {\mathbb K}^\dagger\,{\mathbb D}\,{\mathbb
K}
                   \end{array}\right),
\frac{i}{\sqrt{2}} \left(\begin{array}{ccc}
                     {\mathbb D} & \vline & 0\\
                     \hline
                     0 & \vline & -{\mathbb K}^\dagger\,{\mathbb
D}\,{\mathbb K}
                   \end{array}\right),
i\left(\begin{array}{ccc}
                     0 & \vline & {\mathbb D}\,{\mathbb K}\\
                     \hline
                     0 & \vline & 0           \end{array}\right),
i\left(\begin{array}{ccc}
                     0 & \vline & 0\\
                     \hline
                     {\mathbb K}^\dagger\,{\mathbb D} & \vline & 0
                    \end{array}\right)
             \right],\cr
& &
\label{eq:reps}
\eea
}
where $\mathbb D$ is a real $N/2 \times N/2$ matrix.
One may
furthermore consider quadruplets the entries of which have a definite parity
(the $\mathbb S$'s below stand for scalars and the $\mathbb P$'s for
pseudoscalars)
\begin{equation}
\varphi = ({\mathbb S}^0, \vec {\mathbb P}),
\label{eq:SP}
\end{equation}
and
\begin{equation}
\chi = ({\mathbb P}\,^0, \vec {\mathbb S}).
\label{eq:PS}
\end{equation}
The $\varphi$'s and the $\chi$'s transform alike by the gauge group,
according
to ($i$ and $j$ are $SU(2)$ indices)

\vbox{
\bea
{\mathbb T}^i_L\,.\,{\mathbb M}^j &=& -\frac{i}{2}\left(
              \epsilon_{ijk} {\mathbb M}^k +
                           \delta_{ij} {\mathbb M}^0
                              \right),\cr
{\mathbb T}^i_L\,.\,{\mathbb M}^0 &=&
                                \frac{i}{2}\; {\mathbb M}^i.
\label{eq:actioneven}
\eea
}
The link between the matrices $\mathbb M$ and diquark operators is
straightforwardly established by sandwiching the latter between
$\ol\Psi$ and $\Psi$ and inserting a $\gamma_5$ when needed by  parity.

Because of the CKM rotation, the electroweak
eigenstates $\varphi$  and $\chi$ defined in (\ref{eq:SP},\ref{eq:PS})
are not flavour eigenstates but linear combinations of them.

The flavour or ``strong'' eigenstates are the ones associated with $\mathbb
M$ matrices which have only one nonvanishing entry equal to $1$.

\subsection{Quadratic invariants and electroweak mass scales}
\label{subsec:invariants}

To every quadruplet $({\mathbb M}^0, \vec{\mathbb M})$
is associated a quadratic invariant:
\begin{equation}
{\cal I}
= ({\mathbb M}^0, \vec {\mathbb M})\otimes ({\mathbb M}^0, \vec {\mathbb M})
= {\mathbb {\mathbb M}}\,^0 \otimes {\mathbb {\mathbb M}}\,^0 +
                 \vec {\mathbb M} \otimes \vec {\mathbb M};
\label{eq:invar}
\end{equation}
the ``$\otimes$'' product is a tensor product (not the usual multiplication
of matrices) and means the product of fields as functions of space-time;
$\vec {\mathbb M} \otimes \vec {\mathbb M}$ stands for
$\sum_{i=1,2,3} {\mathbb M}\,^i \otimes  {\mathbb M}\,^i$.

For the relevant cases $N=2,4,6$, there exists a set of $\mathbb D$ real
matrices  such that the algebraic sum
of invariants specified below, extended over all  representations defined by
(\ref{eq:SP},\ref{eq:PS},\ref{eq:reps})
\begin{equation}
{1\over 4}
\left((\sum_{symmetric\ {\mathbb D}'s} - \sum_{antisym\ {\mathbb D}'s})
\left( ({\mathbb S}^0, \vec {\mathbb P})({\mathbb D})
                     \otimes  ({\mathbb S}^0, \vec {\mathbb P})({\mathbb D})
- ({\mathbb P}^0, \vec {\mathbb S})({\mathbb D})
                     \otimes  ({\mathbb P}^0, \vec {\mathbb S})({\mathbb D})
\right)\right)
\label{eq:diaginvar}
\end{equation}
is diagonal both in the electroweak basis and in the basis of flavour
eigenstates. With the coefficient $(1/4)$ chosen in (\ref{eq:diaginvar})
in the electroweak basis,
the normalization in the basis of flavour eigenstates is $(+1/2)$,
with all signs positive.

For the case of two generations, the four $2\times 2$ $\mathbb D$ matrices
($3$ symmetric and $1$ antisymmetric) can be taken as
\begin{equation}
{\mathbb D}_1 = \left( \ba{cc} 1 & 0 \cr
                            0 & 1     \ea \right),\
{\mathbb D}_2 = \left( \ba{rr} 1 & 0 \cr
                            0 & -1    \ea \right),\
{\mathbb D}_3 = \left( \ba{cc} 0 & 1 \cr
                            1 & 0     \ea \right),\
{\mathbb D}_4 = \left( \ba{rr} 0 & 1 \cr
                           -1 & 0     \ea \right).
\label{eq:Dmatrix}
\end{equation}

{From} the property stated above, to each quadruplet can be associated an
arbitrary electroweak mass scale and, for such a choice of
$\mathbb D$ matrices, the degeneracies of electroweak and flavour eigenstates
coincide.

The mass hierarchy of electroweak eigenstates can be made arbitrary without
violating the fundamental symmetries of the theory; it is in particular
disconnected from a hierarchy between quark condensates.

When the chiral and electroweak symmetries are broken, a new splitting can
occur inside each quadruplet between the singlet and the triplet of the
custodial $SU(2)_V$ symmetry studied in \cite{Machet1}; this allows in
particular the splitting between scalar and pseudoscalar mesons and doubles
the number of arbitrary electroweak mass scales up to $N^2$. 

\subsection{The electroweak Lagrangian}
\label{subsec:lagrangian}

The scalar (pseudoscalar) electroweak fields that build up the Lagrangian
are taken to be the ones associated with the set (\ref{eq:Dmatrix}) of matrices
$\mathbb D$ diagonalizing the invariant (\ref{eq:diaginvar}) in both the
electroweak and the flavour basis, and the combination used for the kinetic
terms is the one of (\ref{eq:diaginvar}).

The kinetic terms for the leptons and the gauge fields are the standard
ones.

No Yukawa coupling to quarks is present since they are not fields of the
Lagrangian, and masses are given in a gauge invariant way to the mesons
themselves.

A ``mexican hat'' potential is phenomenologically introduced to trigger the
spontaneous symmetry breaking of the electroweak symmetry.

We report the reader to appendix \ref{app:normalizing} for more details.

\section{$\boldsymbol{K \rightarrow \pi\pi}$ decays: the pure electroweak case}
\label{section:electroweak}
\subsection{The electroweak quadruplets for two generations ($\mathbf N=4$)}
\label{sub:quadruplets}

Specializing to the cas of two generations (we shall not study $CP$
violation effects here), we choose accordingly the four $\mathbb D$ matrices
as in  (\ref{eq:Dmatrix}).

It is useful to write explicitly the four types of multiplets $\Phi({\mathbb
D}_i),i=1\ldots 4$; the one isomorphic to the complex doublet of
the Glashow-Salam-Weinberg model is

\vbox{
\bea
& &\Phi({\mathbb D}_1) = \cr
& & \hskip -2cm \left[
\frac{1}{\sqrt{2}} \left(\ba{rrcrr} 1 &   &\vline &    &    \nonumber\\
                                    & 1 &\vline &    &    \nonumber\\
                                    \hline
                                    &   &\vline &  1 &    \nonumber\\
                                    &   &\vline &    &  1 \ea \right),
\frac{i}{\sqrt{2}} \left(\ba{rrcrr} 1 &   &\vline &   &   \nonumber\\
                                    & 1 &\vline &   &   \nonumber\\
                                    \hline
                                    &   &\vline &-1 &   \nonumber\\
                                    &   &\vline &   & -1   \ea \right),
i \left(\ba{rrcrr}   &  &\vline & c_\theta &  s_\theta \nonumber\\
                             &  &\vline &-s_\theta &  c_\theta \nonumber\\
                            \hline
                             &  &\vline &   &     \nonumber\\
                             &  &\vline &   &  \ea \right),
i \left(\ba{rrcrr}   &   &\vline &   &   \nonumber\\
                             &   &\vline &   &   \nonumber\\
                             \hline
                             c_\theta &-s_\theta &\vline &   &
\nonumber\\
                             s_\theta & c_\theta &\vline &   &   \ea \right)
\right]; \nonumber\\
& &
\label{eq:PHI1}
\eea
}
and the last three are given in Appendix A.
$c_\theta$ and $s_\theta$ stand respectively for the cosine and sine of the
Cabibbo angle $\theta_c$.

We come back later on the normalization (see in particular appendix
\ref{app:normalizing}).

$\varphi({\mathbb D}_1)$ contains one scalar $H = {\mathbb S}^0({\mathbb D}_1)$
and three pseudoscalars $\vec {\mathbb P}({\mathbb D}_1)$. $H$ is the only
scalar matrix with a non-vanishing trace and we take it as the (unique)
Higgs boson.
The  ``mexican hat'' quartic potential that triggers electroweak spontaneous
symmetry breaking is accordingly introduced for this
quadruplet \footnote{and only for this one.}, and the $\vec\mathbb P$'s are the
three associated Goldstone bosons.

The choice of a unique Higgs boson and its interpretation in terms of
$\bar q_i q_j$ operators shows that its getting a non-vanishing
vacuum expectation value (VEV) is equivalent to giving identical non-vanishing
VEV's to all diagonal $(\bar q_i q_i)$ operators, and only to these, building
a bridge between chiral and electroweak symmetry breaking \cite{Machet2}.

Having a single Higgs boson prevents the occurrence of a hierarchy problem
\cite{hierarchy}.
It is also a guarantee not to generate flavour-changing neutral currents
\footnote{
Suppose indeed that, for example, $\la {\mathbb S}^0({\mathbb D}_4)\ra \not
= 0$; then the kinetic terms generate a coupling $\la {\mathbb S}^0({\mathbb
D}_4)\ra\ Z^\mu\ \p_\mu{\mathbb P}^3({\mathbb D}_4)$ between a certain neutral
combination of $K$ and $D$ mesons (see Appendix A) and the $Z$ gauge boson;
the limits on flavour changing neutral currents consequently provide
limits on the VEV's of other eventual Higgs bosons.}
.

Two features of $\varphi({\mathbb D}_1)$ distinguish non-leptonic $K^+$
decays from the ones of neutral kaons:\l
- the components of the three goldstones making the $SU(2)_V$ triplet
$\vec{\mathbb P}({\mathbb D}_1)$ in terms of pseudoscalar mesons
include  the charged $K^\pm$, $D^\pm$ and $\pi^\pm$, the neutral
$\pi^0$ meson and other parts of ``diagonal'' pseudoscalars, but no neutral
$K$ or $D$ meson; so, the latter do not couple to the Higgs boson and a
gauge field while the former do;\l
- the quartic terms in the potential for $\varphi({\mathbb D}_1)$
yields a vertex proportional to $\la H\ra h \otimes \vec{\mathbb P}\otimes
\vec{\mathbb P}$ (see Appendix B) which does not appear for other
quadruplets.\l
They allow one-loop decays for the charged kaons with a Higgs boson
propagating in internal lines, and do not for the neutral kaons.

\subsection{Classical $\boldsymbol{K \rar \pi\pi}$ decays}
\label{sub:classical}

The classical theory is that of a massive gauge theory. We introduce no
gauge-fixing at this level
and choose accordingly the propagators of the gauge bosons to be
\begin{equation}
D^{\mu\nu\ (classical)}_{W,Z}(q)=-i\,\frac{g^{\mu\nu}-q^\mu q^\nu/M_{W,Z}^2}
{q^2-M_{W,Z}^2}.
\label{eq:DWclass}
\end{equation}
There exist non-diagonal couplings between the mesons and the gauge bosons
(see fig.~1 below).
 
The decays $K^+ \rar \pi^+ \pi^0$ and $K_s \rar \pi^+ \pi^-$ are classically
described by the  tree diagrams of fig.~1:

\vbox{
\figskip
\bct
\epsfig{file=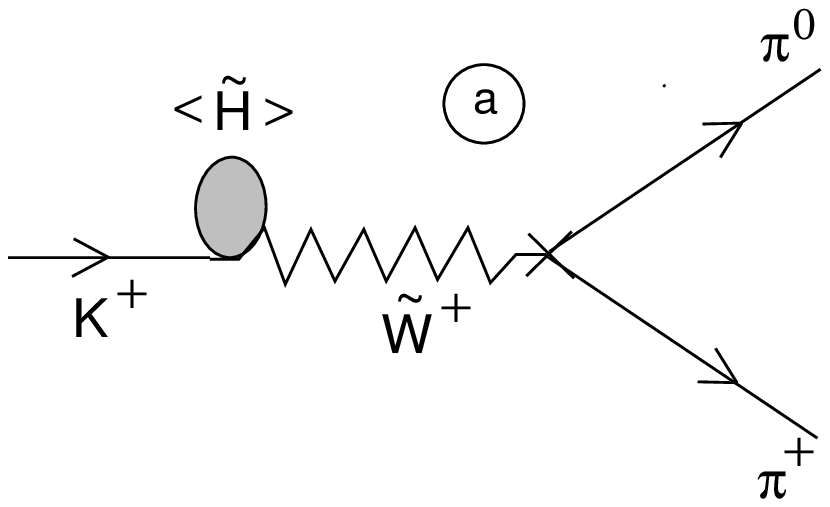,height=4truecm,width=6truecm}
\hskip 2cm
\epsfig{file=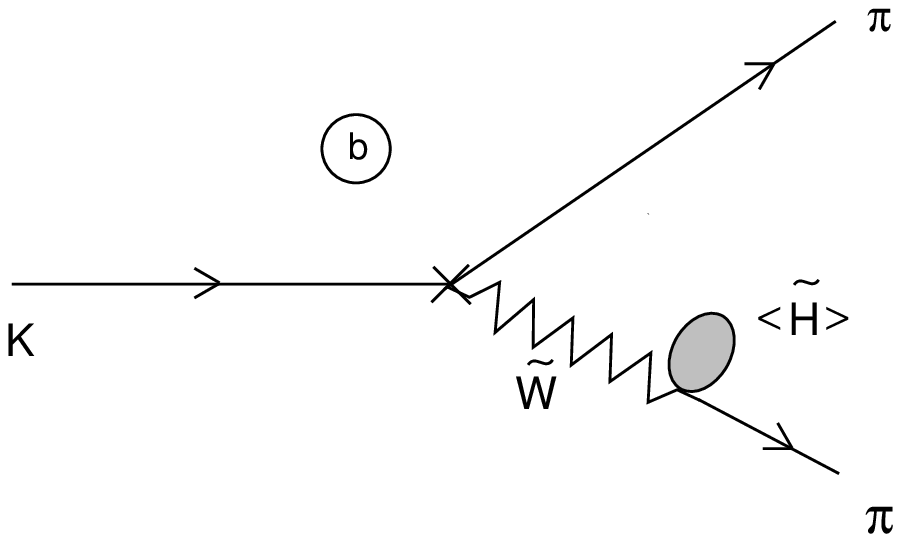,height=5truecm,width=6truecm}
\figskip
{\em Fig.~1: $K^+ \rar \pi^+ \pi^0$ and $K_s \rar \pi^+ \pi^-$ decays at tree
level.}
\ect
}
They are similar to the ones describing leptonic decays of pseudoscalar mesons
(the two leptons coupled there to the gauge fields are replaced here
 by two mesons), which have been used to calculate the normalization
$a$ of the mesonic fields $\mathbb M$  with respect to the observed
mesons: one recovers \cite{Machet1} the usual PCAC result for
\begin{equation}
    a = 2\ \frac{f}{\la H\ra} = 2\sqrt{2}\ \frac{f}{v} =
            f\sqrt{2\sqrt{2}G_F},
\label{eq:a}
\end{equation}
where $f$ is the leptonic decay constant of the mesons (supposed to be the
same for all of them) and $\la H \ra = v/\sqrt{2}$. This entails for
example (see also Appendix \ref{app:normalizing})
\begin{equation}
P^+({\mathbb D}_1) = 2\sqrt{2}\ \frac{f}{v}
         \left(\cos\theta_c(\pi^+ + D_s^+) + \sin\theta_c(K^+ - D^+)\right).
\label{eq:P+}
\end{equation}
We consider that the incoming and outgoing particles are the ``flavour'' or
``strong'' eigenstates $K^+,\pi^\pm,\pi^0$,  and $K_s= (K^0 -
\ol{K^0})/\sqrt{2}$, and the states propagating in internal lines are
electroweak eigenstates (including the massive gauge bosons, the Higgs boson
and the three goldstones).

One has, for example (see also Appendix \ref{subapp:strongew}):\l
\bea
K^+ &\propto& i(\bar u \gamma_5 s) =\frac{1}{2a}\left(
      c_\theta({\mathbb P}^+({\mathbb D}_3) +{\mathbb P}^+({\mathbb D}_4))
  + s_\theta ({\mathbb P}^+({\mathbb D}_1) +{\mathbb P}^+({\mathbb D}_2))
\right),\cr
K_s &\propto& \frac{i}{\sqrt{2}}(\bar d \gamma_5 s - \bar s \gamma_5 d)
= \frac{1}{2a}({\mathbb P}^0 -i{\mathbb P}^3)({\mathbb D}_4).
\label{eq:K+Ks}
\eea
The perturbative series is built from the classical Lagrangian
$\tilde{\cal L}(\tilde\mathbb M)$,
deduced from the one for the $\mathbb M$'s by a
global rescaling by the factor $1/a^2$ (see Appendix \ref{app:normalizing})
\begin{equation}
\tilde{\cal L}(\tilde\mathbb M) \equiv\frac{1}{a^2}{\cal L}({\mathbb M})
= \frac{1}{a^2}{\cal L}(a\tilde{\mathbb M});
\label{eq:scaledL}
\end{equation}
its kinetic terms are diagonal, too, in the basis of strong eigenstates
$\pi, K \ldots$ and start with
\begin{equation}
\tilde{\cal L}(\tilde\mathbb M) =
\frac{1}{2}(D_\mu\pi^0 D^\mu\pi^0 + 2\ D_\mu\pi^+
D^\mu\pi^- + D_\mu K^0 D^\mu\ol{K^0} + 2 D_\mu K^+ D^\mu K^-) +\cdots.
\label{eq:kinetic}
\end{equation}
While both figs.~1a and 1b play a role for $K^+ \rar \pi^+\pi^0$ decays,
fig.~1a vanishes for $K_s \rar \pi^+\pi^-$ because the
$Z$ gauge boson, which would then occur in place of the $W$ boson,
does not couple to neutral kaons (see the footnote in
subsection \ref{sub:quadruplets}).

Because, as it explicitly appears in the expressions for
$D_\mu\pi^+D^\mu\pi^-$ and
$D_\mu K^+D^\mu K^-$, the $Z$ gauge boson does not couple to one charged kaon
and one charged pion
\footnote{Transcribed in the quark language, this corresponds to the absence
of flavour changing neutral current in its customary meaning
 (no coupling of the $Z$ to a $\bar d s$ or $\bar s d$ combination)},
and because $K_s \rar \pi^+\pi^-$ can only be mediated
by a $W^\pm$, only the latter appear in fig.~1b.
 
One gets ($G_F = 1.02\ 10^{-5}/M_{proton}^2$ is the Fermi constant):\l
- for the  amplitudes ($\cal A$ corresponds to fig.~1a and $\cal B$ to
fig.~1b):\l
\begin{equation}
{\cal A}^{tree}_{K^+ \rar \pi^+\pi^0} = \sin\theta_c\cos\theta_c\ 
\sqrt{2} f G_F(m^2_{\pi^+} - m^2_{\pi^0});
\label{eq:Atree}
\end{equation}
\begin{equation}
{\cal B}^{tree}_{K^+ \rar \pi^+\pi^0} = {\cal B}^{tree}_{K_s \rar \pi^+\pi^-}
= - \sin\theta_c\cos\theta_c\ \frac{f}{\sqrt{2}}G_F(m_K^2 - m_{\pi}^2);
\label{eq:Btree}
\end{equation}
- for the decay rates, because $m_{\pi^+}^2 - m_{\pi^0}^2 \ll m_K^2
-m_\pi^2$, we can neglect ${\cal A}$ with respect to ${\cal B}$, and get
accordingly:\l
\bea
\Gamma^{tree}_{K^+ \rar \pi^+\pi^0} \approx\Gamma^{tree}_{K_s \rar \pi^+\pi^-}
&=& \frac{1}{32\pi}\sin^2\theta_c\cos^2\theta_c\,G_F^2\,\frac{f^2}{m_K^2}
(m_K^2 - m_\pi^2)^2 \sqrt{m_K^2 - 4 m_\pi^2}\cr
&\approx& 9.5\ 10^{-17} GeV,
\label{eq:Gammatree}
\eea
which is to be compared with the experimental values:
\begin{equation}
\Gamma^{exp}_{K^+ \rar \pi^+\pi^0} = 1.1\ 10^{-17} GeV <
\Gamma^{tree}_{\overset{K^+ \rar \pi^+\pi^0}{\ K_s \rar \pi^+\pi^-}} <
\Gamma^{exp}_{K_s \rar \pi^+\pi^-} = 5.125\ 10^{-15} GeV.
\label{eq:Gammaexp}
\end{equation}
%
\subsubsection{The case of $\boldsymbol{K_s \rar \pi^0\pi^0}$ decays}
\label{subsub:pi0pi0classical}

There exist no electroweak contribution to $K_s \rar \pi^0\pi^0$ at the tree
level.
This is the first sign of a drastic violation of the strong isospin symmetry
by electroweak interactions.

The tree diagrams are vanishing because;\l
- there is no coupling of $K_s = (1/2)({\mathbb P}^0 - i{\mathbb P}^3)({\mathbb
D}_4)$ to the $Z$ gauge boson (see the footnote in subsection
\ref{sub:quadruplets}); this eliminates the diagrams of fig.~1a;\l
- the $Z$ boson does not couple to $K_s$ and $\pi^0$, because
${\mathbb T}^3_L$ acting on $K_s$ yields ${\mathbb S}^0({\mathbb D}_4)$ or
${\mathbb S}^3({\mathbb D}_4)$ which do not contain $\pi^0$ (see Appendix
\ref{subapp:ewstrong}), and ${\mathbb
T}^3_L$ acting on $\pi^0$ does not yield any $K_s$; this eliminates the
diagrams of the type of fig.~1b.

\subsubsection{Summary of the classical electroweak contributions}
\label{subsub:classicsum}

The classical electroweak theory faces three problems:\l
- it gives too large a decay rate (by a factor $8.6$) for $K^+ \rar
\pi^+\pi^0$;\l
- it gives too small a decay rate (by a factor $54$) for $K_s \rar
\pi^+\pi^-$;\l
- $K_s \rar \pi^0\pi^0$ decays do not occur.

\subsection{The electroweak one-loop corrections to
$\boldsymbol{K^+ \rar \pi^+ \pi^0}$}
\label{sub:ew1loop}

\subsubsection{The loop expansion}
\label{subsub:expans}

The global normalization of the Lagrangian (see eq.~(\ref{eq:scaledL}) and
Appendix \ref{app:normalizing}) plays a role at the quantum level.
Let us indeed consider its influence on the generating functional $Z$
\begin{equation}
Z = e^{\frac{i}{\hbar}\int d^4x {\cal L}({\mathbb M}(x))}
  = e^{a^2\frac{i}{\hbar}\int d^4x \frac{1}{a^2}{\cal L}(a\tilde{\mathbb M}(x))}
  = e^{a^2\frac{i}{\hbar}\int d^4x \tilde{\cal L}(\tilde{\mathbb M}(x))}.
\label{eq:Z}\end{equation}
The first equality is a trivial identity, where we just expressed the
generic fields $\mathbb M$ in terms of the rescaled one $\tilde{\mathbb M}$;
the second equality is only the definition of $\tilde{\cal L}$.
Eq.~(\ref{eq:Z}) shows that while the parameter driving the loop expansion for
${\cal L}(\mathbb M)$ is $\hbar$, it is $\hbar/a^2$ for $\tilde{\cal
L}(\tilde{\mathbb M})$.

The consequence of the rescaling of the fields and of the Lagrangian
is accordingly that, for the same number of external legs, the one-loop
amplitude for a given process gets an extra $1/a^2  = v^2/8f^2$ factor with
respect to the tree amplitude.

\subsubsection{1-loop diagrams for $\boldsymbol{K^+ \rar \pi^+ \pi^0}$ decays}
\label{subsub:ewK+pi+pi0}

The only six non-vanishing diagrams are drawn in fig.~2. In Appendix C we
display other diagrams and explain why they identically vanish.

\vbox{
\figskip
\epsfig{file=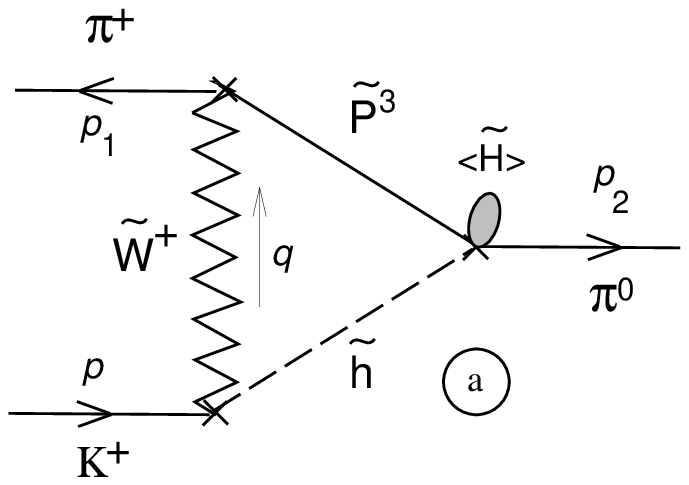,height=4truecm,width=5truecm}
\epsfig{file=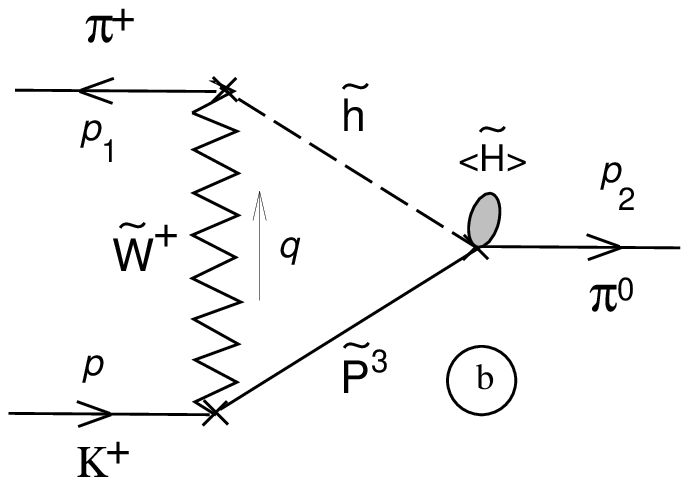,height=4truecm,width=5truecm}
\epsfig{file=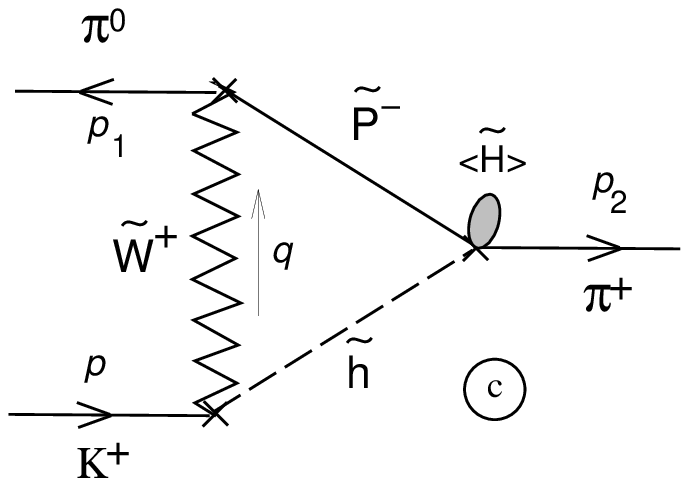,height=4truecm,width=5truecm}
\figskip
\epsfig{file=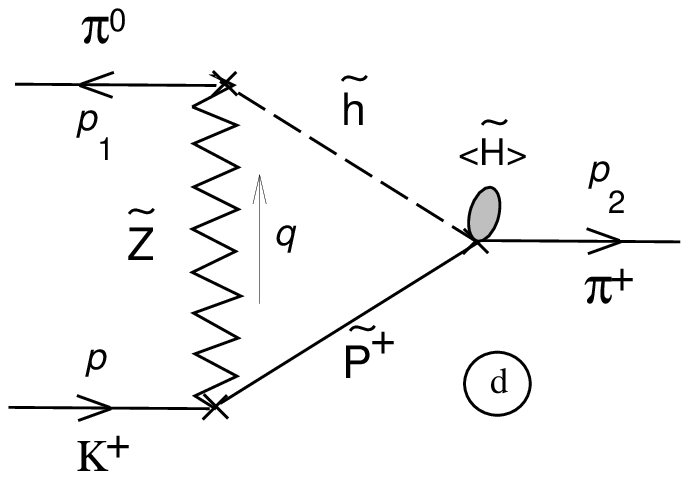,height=4truecm,width=5truecm}
\epsfig{file=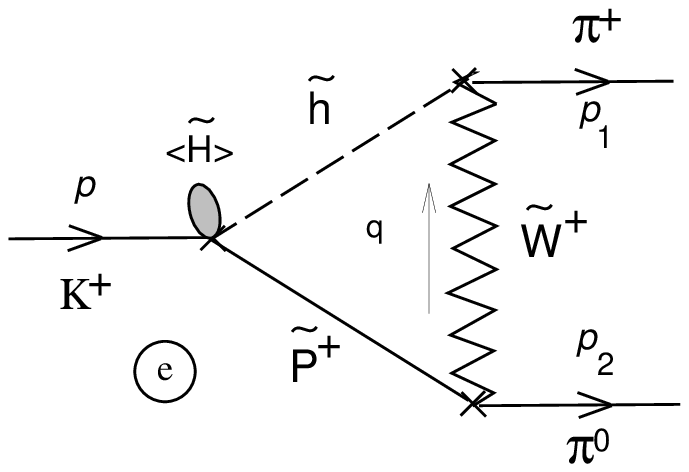,height=4truecm,width=5truecm}
\epsfig{file=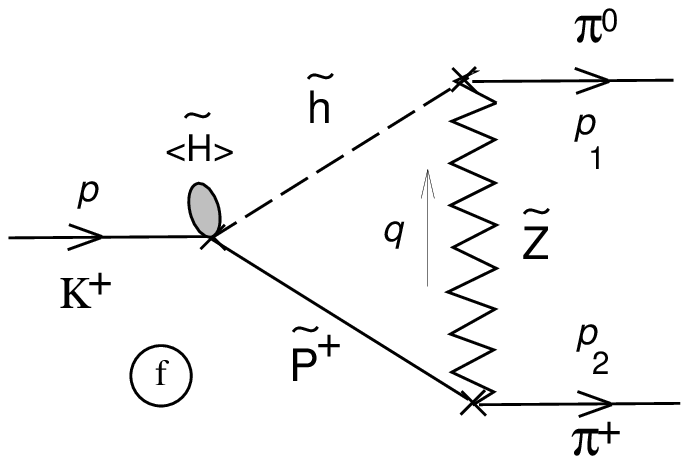,height=4truecm,width=5truecm}
\figskip
\bct
{\em Fig.~2: nonvanishing 1-loop electroweak diagrams for
$K^+ \rar \pi^+ \pi^0$.}
\ect
}
Because all diagrams involve the trilinear coupling
between the Higgs and two pseudoscalar mesons characteristic of the
quadruplet $\varphi({\mathbb D}_1)$, the latter can only
be the Goldstone bosons of the quadruplet $\varphi({\mathbb D}_1)$.

\subsubsection{1-loop diagrams for $\mathbf{K_s \rar \pi\pi}$ decays}
\label{subsub:ewKspipi}

All are vanishing because;\l
- $K_s$ is linked to $\varphi({\mathbb D}_4)$ and to $\chi({\mathbb D}_4)$
(see eqs.~(\ref{eq:K+Ks},\ref{eq:SP},\ref{eq:PS}))
and we supposed that the only field with a non-vanishing VEV is ${\mathbb
S}^0({\mathbb D}_1)$; so, there is no trilinear coupling with which one-loop
diagrams could be built; this eliminates the diagrams of fig.~2 and
fig.~10a;\l
- the other diagrams of the types described in figs.~10b,c vanish for the same
reasons as stated in Appendix C.

\subsubsection{Explicit computations}
\label{subsub:ewcomput}

We perform the computations in the Landau gauge; it is
coherent with considering both the three massive gauge bosons {\em and}
the three goldstones in internal lines; indeed, while a massive gauge boson
has three degrees of freedom, a transverse one has only two.
In this gauge, all logarithmic divergences cancel and the results are
finite.

One finds the following results for the amplitudes ${\cal A}^{a,b,c,d,e,f}$
corresponding respectively to the diagrams of fig.~2a,b,c,d,e,f.:

\hskip -1cm
\vbox{
\bea
{\cal A}^a + {\cal A}^b &=&  0,\cr
{\cal A}^b + {\cal A}^d &=& 4i\ s_\theta c_\theta\ f\ G_F^2\ M_W^2 m_H^2 \cr
&&\left(
I_1(M_W^2, m_\pi^2, m_K^2) - \frac{1}{c_W^2} I_1(M_Z^2, m_\pi^2, m_K^2)
-I_2(M_W^2, m_\pi^2, m_K^2) - \frac{1}{c_W^2} I_2(M_Z^2, m_\pi^2, m_K^2)
\right),\cr
{\cal A}^e + {\cal A}^f &=& -({\cal A}^b + {\cal A}^d)(p \leftrightarrow
-p_2),
\label{eq:A}
\eea
}

where $p$ and $p_2$ are respectively the momenta of the incoming kaon and of
one outgoing pion as displayed in figs.~2; $I_1$ and $I_2$
are dimensionless convergent $1$-loop integrals respectively given by:
\bea
I_1(M^2,  m_\pi^2, m_K^2) &=& p.p_1\int \frac{d^4 q}{(2\pi)^4}\ 
                \frac{1}{(p_1 -q)^2 -m_H^2}\ 
                \frac{1}{(p-q)^2}\ 
                \frac{1}{q^2 - M^2},\cr
I_2(M^2,  m_\pi^2, m_K^2) &=& \int \frac{d^4 q}{(2\pi)^4}\ (q.p)\,(q.p_1)\ 
                \frac{1}{(p_1 -q)^2 -m_H^2}\ 
                \frac{1}{(p-q)^2}\ 
                \frac{1}{q^2 - M^2}\ 
                \frac{1}{q^2}.
\label{eq:I1I2}
\eea
Their explicit analytic expressions  are given in Appendix D.

$c_W$ and $s_W$ stand respectively for $\cos\theta_W$ and $\sin\theta_W$,
where $\theta_W$ is the Weinberg angle.

Using $p.p_1 = (1/2)m_K^2$ and $p.p_1+p_1.p_2 = m_K^2 - m_\pi^2$, one gets
\begin{equation}
{\cal A}^{1\ loop}_{K^+\rar\pi^+\pi^0}= -\frac{3}{16\pi^2}
\ \sin\theta_c \cos\theta_c\ G_F^2\ f\ m_H^2\ (m_K^2 - m_\pi^2)
\left(
\frac{M_Z^2}{M_Z^2 - m_H^2}\ \ln\frac{M_Z^2}{m_H^2}
-\frac{M_W^2}{M_W^2 - m_H^2}\ \ln\frac{M_W^2}{m_H^2}
\right),
\label{eq:A1loop}
\end{equation}
which exhibits a cancelation between the $W$ and $Z$ contributions.

It stays finite when $m_H^2 \rar M_W^2, M_Z^2$
and, when $m_H^2 \rar \infty$

\hskip -1.5cm
\vbox{
\begin{equation}
{\cal A}^{1\ loop}_{K^+\rar\pi^+\pi^0}
\underset{m_H^2 \rar \infty}{\approx}
-\frac{3}{16\pi^2}\ \sin\theta_c \cos\theta_c\ G_F^2\ f\ (m_K^2 - m_\pi^2)
\left( M_W^2\ \ln M_W^2 - M_Z^2\ \ln M_Z^2 +(M_Z^2 - M_W^2)\ \ln{m_H^2}\right).
\label{eq:Ainfty}
\end{equation}
}
%
\paragraph{The Higgs contribution to $\boldsymbol{K^+ \rar \pi^+\pi^0}$ is
negligeable.}

To have an idea of the importance of the one-loop contributions, we consider
the ratio $\cal R$ of the 1-loop amplitude (\ref{eq:A1loop})
and the tree amplitude (\ref{eq:Btree}); it can be of course argued that the
former, unlike the latter, has been computed without gauge fixing,
but one considers it here only as a rough estimate of
the experimental value (see eq.~(\ref{eq:Gammaexp})); $\cal R$ is given by:
\begin{equation}
{\cal R} = \frac{3\sqrt{2}}{16\pi^2}G_F m_H^2
\left(
\frac{M_Z^2}{M_Z^2 - m_H^2}\ \ln\frac{M_Z^2}{m_H^2}
-\frac{M_W^2}{M_W^2 - m_H^2}\ \ln\frac{M_W^2}{m_H^2}
\right),
\label{eq:R}
\end{equation}
and is plotted of fig.~3 for a large range of values of $m_H \in
[0,100\,TeV]$.

One concludes that the Higgs boson can play no role in explaining the
$\Delta I =1/2$ rule for $K\rar \pi\pi$ decays
 when only electroweak interactions are taken into account.

\vbox{
\figskip
\bct
\epsfig{file=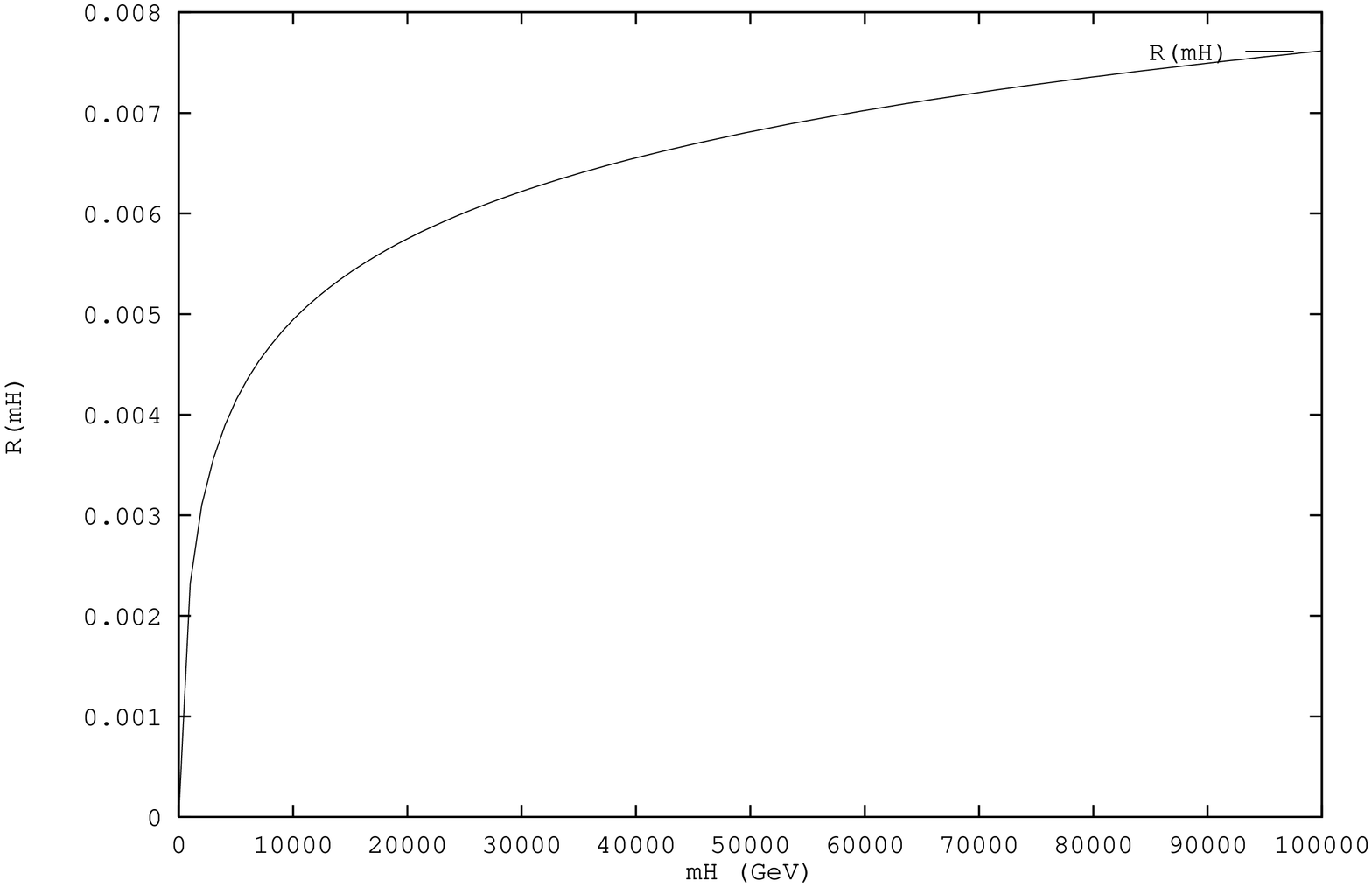,height=15truecm,width=8truecm,angle=-90}
\figskip
{\em Fig.~3: the ratio of one-loop over tree electroweak amplitudes
for $K^+ \rar \pi^+\pi^0$.}
\ect
}
%
\subsection{Conclusion}
\label{sub:exconclusion}

$SU(2)_L\times U(1)$ electroweak interactions for mesons transforming like
$(\bar q_i q_j)$ or $(\bar q_i \gamma_5 q_j)$ operators cannot account
for $K_s \rar \pi^0\pi^0$ decays nor
for the damping of $K^+ \rar \pi^+\pi^0$ with respect to $K_s \rar \pi^+\pi^-$.
Another type of interactions is consequently needed to account for the
observed $K \rar \pi\pi$ transitions.

\section{$\boldsymbol{K \rightarrow \pi\pi}$ decays: introducing another
interaction between mesons}
\label{section:newint}

Would this be the only problem, the absence of a $\pi^0\pi^0$ final state
in the electroweak decays of $K_s$  could be cured by invoking final state
strong interactions in the process $\pi^+\pi^- \leftrightarrow \pi^0\pi^0$;
in the quark language, this is tantamount to a simple
reshuffling of the quark lines.

However, this cannot explain the large effect of isospin breaking
between $K^+ \rar \pi^+\pi^0$ and $K_s \rar \pi^+\pi^-$ decays.

This is why we shall adopt another point of view and introduce another
interaction between mesons, which in particular respects the two
symmetries generally attributed to strong interactions: flavour symmetry
and parity conservation.
Its interpretation is again clearer in the quark picture.

Consider an incoming mesonic state built from the two quarks $q_i$ and
$q_j$, $(\bar q_i q_j)$ for a scalar and $(\bar q_i \gamma_5 q_j)$ for a
pseudoscalar.

Since chiral symmetry is broken (supposedly by strong interactions),
the vacuum expectation values of diagonal
scalar diquark operators are nonvanishing $\la \bar q_\alpha q_\alpha \ra
\not = 0$. This means that in the true vacuum, diagonal quark pairs can be
freely created. Because of the flavour symmetry of strong interactions and
because they conserve parity, the
only combination of quark pairs that can occur is the flavour diagonal
combination $(\bar u u + \bar c c + \bar d d + \bar s s)$ (which corresponds
to the Higgs boson).

We then invoke a reshuffling of the quark lines, which induces a trilinear
coupling between mesons depicted in fig.~4, in which the dark
square represents the (unknown) $\Lambda$-coupling.

For example we can have the coupling (in the quark language)
\bea
(\bar d \gamma_5 c)\  \underset{\Lambda}{\longrightarrow}
&&(\bar d u) (\bar u \gamma_5 c) + (\bar d \gamma_5 u) (\bar u c) +
(\bar d c) (\bar c \gamma_5 c) + (\bar d \gamma_5 c) (\bar c c)\cr
+&&(\bar d d) (\bar d \gamma_5 c) + (\bar d \gamma_5 d) (\bar d c) +
(\bar d s) (\bar s \gamma_5 c) + (\bar d \gamma_5 s) (\bar s c).
\label{eq:Lambdaexample}
\eea

\vbox{
\figskip
\bct
\epsfig{file=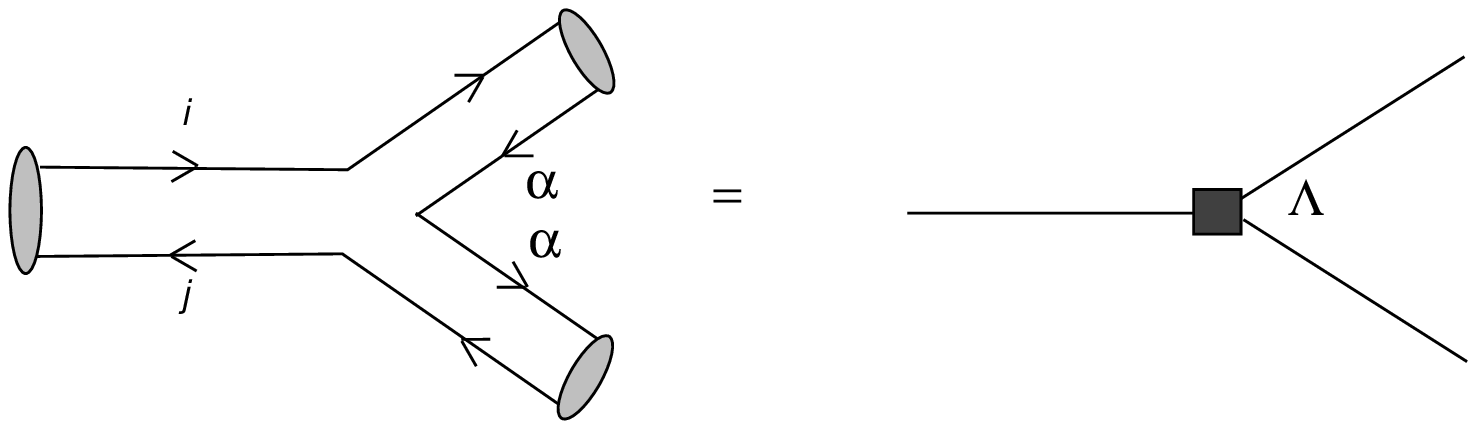,height=4truecm,width=12truecm}
\figskip
{\em Fig.~4: trilinear meson coupling induced by $\Lambda$-interactions.}
\ect
}
$\Lambda$ has the dimension of  $[mass]$ (see also eq.~(\ref{eq:Aint}) below).
As all scalar and pseudoscalar fields have dimension $[mass]$, the new
coupling does not harm renormalizability.

The trilinear coupling violates electroweak symmetries and mix components of
different quadruplets
\footnote{It also yields, when one of the external legs is the Higgs boson,
identical mass terms for all scalar and pseudoscalar
mesons through its condensation.}.
Since all quadruplets are now concerned,
the relevant diagrams, which all now occur at the one-loop level,
are more numerous than in the pure electroweak case. 
The diagrams which can play a role in $K^+ \rar \pi^+\pi^0$ decays are
displayed in Appendix E, those for $K_s \rar \pi\pi$ in Appendix F.

We shall make the following simplifying hypothesis in the computations of
the diagrams which now mix electroweak and $\Lambda$-interactions:\l
- {\em all internal lines are electroweak eigenstates};\l
- {\em they furthermore correspond to the eight quadruplets
associated with the matrices ${\mathbb D}_{1=1\ldots 4}$ of
eq.~(\ref{eq:Dmatrix}), which
diagonalize the kinetic terms both in the electroweak and in the strong
basis.}

Each diagram is computed as the sum of the amplitudes corresponding to each
given intermediate electroweak eigenstate.

The second part of the above assumption is
not guaranteed (except for $\Phi({\mathbb D}_1)$ since it includes
the Higgs boson and the three goldstones (see eq.~(\ref{eq:PHI1})),
which are always electroweak mass eigenstates).

Each external line  is a linear combination of electroweak eigenstates (see
Appendix \ref{subapp:strongew}).

That the two final states must be pseudoscalars eliminates several potential
diagrams (see Appendix \ref{subapp:vanK+pi+pi0}); also, even in the case
where the two outgoing states are
pseudoscalars, they may have a vanishing projection on the expected
two-pions states (see Appendix \ref{subapp:strongew}),
which eliminates other diagrams.

\subsection{Explicit computations: an example}
\label{sub:example}

We outline below the main steps of the computation of the diagram
corresponding to fig.~11$\alpha$ for the decay $K^+ \rar \pi^+\pi^0$; the first
intermediate electroweak state is here a scalar ${\mathbb S}^0$.

The ingoing $K^+$ meson projects on electroweak eigenstates according to
eq.~(\ref{eq:K+Ks}).
 
The coupling with the $W$ gauge boson connects any ${\mathbb P}^+$ to the
${\mathbb S}^0$ {\em of the same quadruplet}.

The electroweak coupling between ${\mathbb P}^+$, ${\mathbb S}^0$ and $W$
being the same for all quadruplets, the combination of electroweak
eigenstates that enters the $\Lambda$-vertex  is the same linear
combination as the one occurring in eq.~(\ref{eq:K+Ks})
$(1/2a)\ \left(
s_\theta({\mathbb S}^0({\mathbb D}_1) + {\mathbb S}^0({\mathbb D}_2))
+c_\theta ({\mathbb S}^0({\mathbb D}_3) + {\mathbb S}^0({\mathbb D}_4))
\right) \propto s_\theta(\bar u u + \bar s s) +c_\theta(\bar u c + \bar d
s)$, where we have used the quark notation in the last equality.

Wishing eventually to account for the mass splittings between
intermediate states, we treat separately the amplitude corresponding
to any given ${\mathbb S}^0({\mathbb D}_i)$.

Let us consider for example the amplitude corresponding to ${\mathbb
S}^0({\mathbb D}_4) \propto ((\bar u c) - (\bar c u) + (\bar d s) - (\bar s d))$
as the first intermediate state.

By the model of $\Lambda$-interactions proposed above, this scalar combination
yields

\vbox{
\bea
&& (\bar u c) - (\bar c u) + (\bar d s) - (\bar s d)
\underset{\Lambda}{\longrightarrow}
\cr
&& ((\bar u u) (\bar u c) + (\bar u c) (\bar c c) +
         (\bar u d) (\bar d c) + (\bar u s) (\bar s c)
   -(\bar c u) (\bar u u) - (\bar c c) (\bar c u)
       - (\bar c d) (\bar d u) - (\bar c s) (\bar s u)\cr
 && +(\bar d u) (\bar u s) + (\bar d c) (\bar c s) +
         (\bar d d) (\bar d s) + (\bar d s) (\bar s s)
    - (\bar s u) (\bar u d) - (\bar s c) (\bar c d) -
        (\bar s d) (\bar d d) - (\bar s s) (\bar s d))\cr
&+&
((\bar u \gamma_5 u) (\bar u \gamma_5 c) 
       + (\bar u \gamma_5 c) (\bar c \gamma_5 c)
       + (\bar u \gamma_5 d) (\bar d \gamma_5 c) 
       + (\bar u \gamma_5 s) (\bar s \gamma_5 c)\cr
       && - (\bar c \gamma_5 u) (\bar u \gamma_5 u)
       - (\bar c \gamma_5 c) (\bar c \gamma_5 u)
       - (\bar c \gamma_5 d) (\bar d \gamma_5 u)
       - (\bar c \gamma_5 s) (\bar s \gamma_5 u))\cr
&+& ((\bar d \gamma_5 u) (\bar u \gamma_5 s)
       + (\bar d \gamma_5 c) (\bar c \gamma_5 s)
       + (\bar d \gamma_5 d) (\bar d \gamma_5 s)
       + (\bar d \gamma_5 s) (\bar s \gamma_5 s)\cr
       && - (\bar s \gamma_5 u) (\bar u \gamma_5 d)
       - (\bar s \gamma_5 c) (\bar c \gamma_5 d)
       - (\bar s \gamma_5 d) (\bar d \gamma_5 d)
       - (\bar s \gamma_5 s) (\bar s \gamma_5 d)).
\label{eq:Lambdaint}
\eea
}

Of the two states outgoing the strong vertex, only one undergoes again
electroweak interactions; the other one is a final outgoing state.

As the two final states must be pseudoscalars, all terms in
eq.~(\ref{eq:Lambdaint}) involving two scalars can be discarded since,
otherwise, a scalar will remain among the final states.

As the two final states have strangeness and charm zero, all contributions
involving two ``$s$'' quarks, or two ``$c$'' quarks or one ``$s$'' and one
``$c$'' can be discarded since at least one of them would remain in the
final states.

We can thus restrict eq.~(\ref{eq:Lambdaint}) above to

\vbox{
\bea
(\bar u c) - (\bar c u) + (\bar d s) - (\bar s d)&&
\underset{\Lambda}{\longrightarrow}
(\bar u \gamma_5 u) (\bar u \gamma_5 c) 
       + (\bar u \gamma_5 d) (\bar d \gamma_5 c)
      - (\bar c \gamma_5 u) (\bar u \gamma_5 u)
      - (\bar c \gamma_5 d) (\bar d \gamma_5 u)\cr
+&&(\bar d \gamma_5 u) (\bar u \gamma_5 s)
       + (\bar d \gamma_5 d) (\bar d \gamma_5 s)
        - (\bar s \gamma_5 u) (\bar u \gamma_5 d)
         - (\bar s \gamma_5 d) (\bar d \gamma_5 d)
       + \cdots.\cr
&&
\label{eq:truncation}
\eea
}
The final states being strangeless and charmless, the vertex where
the $W^+$ gauge boson annihilates must
operate on the strange or charmed intermediate state if there exists any,
so as to transform it into a charmless and strangeless state.
This state must furthermore be neutral or negatively charged. It can thus
only be $(\bar d \gamma_5 c)$, $(\bar c \gamma_5 u)$, $(\bar d \gamma_5 s)$ or
$(\bar s \gamma_5 d)$,
and we can forget about the term $(\bar c \gamma_5 d) (\bar d \gamma_5 u)$
and $(\bar d \gamma_5 u) (\bar u \gamma_5 s)$ 
in eq.~(\ref{eq:truncation}): the incoming $W^+$ could only connect to
 $(\bar d \gamma_5 u)$
such that  $(\bar u \gamma_5 s)$ and $(\bar c \gamma_5 d)$ would be final
states which strangeness or charm, which cannot correspond to a pion.

The electroweak vertex where the $W^+$ is annihilated
can only concern the intermediate states $(\bar u \gamma_5 c)$ and $(\bar c
\gamma_5 u)$ (because they are charmed and neutral), $(\bar d \gamma_5 s)$ and
$(\bar s \gamma_5 d)$ (because they are strange and neutral), $(\bar d \gamma_5
c)$ and $(\bar s \gamma_5 u)$ because they are charmed or strange and have a
negative charge.
These we expand in terms of electroweak eigenstates, using the formul\ae  of
Appendix \ref{subapp:strongew}.
A ${\mathbb P}^0$ intermediate state can only be connected by a weak generator
${\mathbb T}_L^-$
to a ${\mathbb S}^-$, that is the creation of a {\em scalar} with charge $+1$;
it cannot correspond to a pion; thus in the expansion (\ref{eq:truncation})
 above, all ${\mathbb P}^0$ intermediate states can be discarded and one
has only to consider
${\mathbb P}^3$ or  ${\mathbb P}^-$ as possible electroweak intermediate
states.

The states outgoing the strong vertex which are not acted upon by electroweak
interactions, already expressed in terms of strong eigenstates, have just to
be projected on pion states; omitting the $a$ factors, one has for example
(see Appendix \ref{subapp:strongew}):

\vbox{
\bea
(\bar u \gamma_5 d) &\propto& -i\ \pi^+; \cr
(\bar u \gamma_5 u) &\propto& -i\ \frac{1}{\sqrt{2}}(\pi^0 + \eta);\cr
(\bar d \gamma_5 d) &\propto& -i\ \frac{1}{\sqrt{2}}(\eta - \pi^0),
\label{eq:projection}
\eea
}
and the terms containing $\eta$ are to be discarded.

Likewise, the electroweak final states coming out of the interaction vertex
have to be projected on the pion states, according with the formulae  of
Appendix \ref{subapp:ewstrong}.

One of course only keeps in the global amplitude terms proportional to
$\sin\theta_c \cos\theta_c$ and drops higher powers of $\sin\theta_c$.

Collecting all the factors and in particular taking into account  the
appropriate normalization factors $a$ for the fields and coupling constants,
one gets the following expression
\bea
{\cal S}_\alpha^{+0}({\mathbb S}^0({\mathbb D}_4),W)&=&
 s_\theta c_\theta  \Lambda \frac{ag^2}{32\sqrt{2}}
\int \frac{d^4 q}{(2\pi)^4}(2p-q)^\mu (2p_1-q)^\nu D_{\mu\nu}^W(q)
                       D_{{\mathbb S}^0({\mathbb D}_4)}(p-q)\cr
&&(D_{{\mathbb P}^+({\mathbb D}_1)} -
                                 D_{{\mathbb P}^3({\mathbb D}_4)})(p_1-q).\cr
&&
\label{eq:S+04}
\eea
The $D$'s stand for the propagators of the corresponding intermediate states.

The same procedure must be repeated for ${\mathbb S}^0({\mathbb D}_{1,2,3})$
in the first internal line.
One finds:
\bea
{\cal S}_\alpha^{+0}({\mathbb S}^0({\mathbb D}_1),W)&=&
2 s_\theta c_\theta  \Lambda \frac{ag^2}{32\sqrt{2}}
\int \frac{d^4 q}{(2\pi)^4}(2p-q)^\mu (2p_1-q)^\nu D_{\mu\nu}^W(q)
                       D_{{\mathbb S}^0({\mathbb D}_1)}(p-q)\cr
&&(D_{{\mathbb P}^+({\mathbb D}_1)} + D_{{\mathbb P}^+({\mathbb D}_2)}
- D_{{\mathbb P}^3({\mathbb D}_1)}- D_{{\mathbb P}^3({\mathbb D}_2)})(p_1-q);\cr
&& \cr
{\cal S}_\alpha^{+0}({\mathbb S}^0({\mathbb D}_2),W)&=&
2 s_\theta c_\theta  \Lambda \frac{ag^2}{32\sqrt{2}}
\int \frac{d^4 q}{(2\pi)^4}(2p-q)^\mu (2p_1-q)^\nu D_{\mu\nu}^W(q)
                       D_{{\mathbb S}^0({\mathbb D}_2)}(p-q)\cr
&&(D_{{\mathbb P}^+({\mathbb D}_1)} + D_{{\mathbb P}^+({\mathbb D}_2)}
- D_{{\mathbb P}^3({\mathbb D}_1)}- D_{{\mathbb P}^3({\mathbb
D}_2)})(p_1-q);\cr
&& \cr
{\cal S}_\alpha^{+0}({\mathbb S}^0({\mathbb D}_3),W)&=&
 s_\theta c_\theta  \Lambda \frac{ag^2}{32\sqrt{2}}
\int \frac{d^4 q}{(2\pi)^4}(2p-q)^\mu (2p_1-q)^\nu D_{\mu\nu}^W(q)
                       D_{{\mathbb S}^0({\mathbb D}_3)}(p-q)\cr
&&(D_{{\mathbb P}^3({\mathbb D}_1)} + D_{{\mathbb P}^3({\mathbb D}_3)}
- D_{{\mathbb P}^+({\mathbb D}_1)}- D_{{\mathbb P}^+({\mathbb D}_3)})(p_1-q).\cr
&&
\label{eq:S+0123}
\eea
The amplitude for the diagram of fig.~11$\alpha$ is the sum
\begin{equation}
{\cal S}_\alpha^{+0}(W) = {\cal S}_\alpha^{+0}({\mathbb S}^0({\mathbb D}_1),W)
                     + {\cal S}_\alpha^{+0}({\mathbb S}^0({\mathbb D}_2),W)
                     + {\cal S}_\alpha^{+0}({\mathbb S}^0({\mathbb D}_3),W)
                     + {\cal S}_\alpha^{+0}({\mathbb S}^0({\mathbb D}_4),W).
\label{eq:S+0W}
\end{equation}
${\cal S}_\alpha^{+0}({\mathbb S}^0({\mathbb D}_1),W)$, 
${\cal S}_\alpha^{+0}({\mathbb S}^0({\mathbb D}_2),W)$ and
${\cal S}_\alpha^{+0}({\mathbb S}^0({\mathbb D}_3),W)$
vanish as soon as the three pseudoscalars inside each triplet
$\vec{\mathbb P}({\mathbb D}_1)$, $\vec{\mathbb P}({\mathbb D}_2)$,
$\vec{\mathbb P}({\mathbb D}_3)$ are degenerate. We suppose that this is the
case, so as to conserve in particular the custodial $SU(2)_V$ symmetry
\cite{Machet1}, and approximate
\begin{equation}
{\cal S}_\alpha^{+0}(W)
\approx {\cal S}_\alpha^{+0}({\mathbb S}^0({\mathbb D}_4),W).
\label{eq:S+0approx}
\end{equation}
This diagram is in particular a function of the Higgs
($\equiv {\mathbb S}^0({\mathbb D}_1)$) mass and of the mass difference
between $\vec{\mathbb P}({\mathbb D}_4)$ and the three goldstones
$\vec{\mathbb P}({\mathbb D}_1)$.

\subsection{Results for $\boldsymbol{K \rar \pi \pi}$ amplitudes}
\label{sub:general}

The computations of all relevant diagrams go along the same lines as
explained above.

\subsubsection{General analytic results}
\label{subsub:analytic}

They are written in Appendix \ref{app:K+pi+pi0amps} for $K^+ \rar \pi^+ \pi^0$,
in Appendix \ref{app:Kspi+pi-amps} for $K_s \rar \pi^+\pi^-$ and in
Appendix \ref{app:Kspi0pi0amps} for $K_s \rar \pi^0 \pi^0$.

\subsubsection{Approximations}
\label{subsub:approx}

We choose, for the same reasons as in section \ref{subsub:ewcomput},
to work in the Landau gauge: both the three massive gauge fields and the
three goldstones are taken into account in the internal lines.
In this gauge, the (purely electroweak) tree amplitudes vanish, and all the
loop amplitudes that we compute are again finite.

As there are {\em a priori} two possible mass scales per electroweak
quadruplet
if the scheme of symmetry breaking preserves the $SU(2)_V$ custodial symmetry,
the masses in internal lines can take sixteen different values.
This general case being uncontrollable, one must work within certain
approximations. This is all the more legitimate as we already made an
hypothesis on the nature of the internal lines and because the goal of this
study is not to derive the spectrum of electroweak
mass eigenstates from the $K \rar \pi\pi$ amplitudes, but instead to show
how large deviations from the strong isospin symmetry can occur.

We shall first work in the symmetric limit where all scalar and
pseudoscalars are degenerate with mass $m$;
we shall then study the case when there are two mass scales, one for the
scalars, $m_S$, and one for the pseudoscalars, $m_P$;
we shall finally evoke the more realistic case in which there are four
mass scales, the vanishing one, $m_G$, for the three goldstones of the broken
electroweak symmetry, a second one, $m_P$, for the other pseudoscalars,
a third one, $m_S$, for all scalars but the Higgs boson which is given
a fourth one, $m_H$.

Accordingly, the amplitudes under concern write:\l
$\bullet$
\begin{equation}
{\cal S}_{K^+ \rar \pi^+\pi^0}={\cal S}^W_{K^+ \rar \pi^+\pi^0}+
                               {\cal S}^Z_{K^+ \rar \pi^+\pi^0},
\label{eq:SK+pi+pi0}
\end{equation}
with
\bea
{\cal S}^W_{K^+ \rar \pi^+\pi^0}&=&
              i s_\theta c_\theta f M_W^2 \sqrt{2}G_F A \left(
        -3(J1-J2)(M_W,m_H,m_G) -5(J1-J2)(M_W,m_S,m_P)\right),\cr
{\cal S}^Z_{K^+ \rar \pi^+\pi^0}&=&
              \frac{i}{c_W^2}s_\theta c_\theta f M_W^2 \sqrt{2}G_F A \left(
        (J1-J2)(M_Z,m_H,m_P) +3 (J1-J2)(M_Z,m_H,m_G)\right.\cr
       &&\left.\hskip 2cm
+(J1-J2)(M_Z,m_S,m_P)+3(J1-J2)(M_Z,m_S,m_G)\right);
\label{eq:SK+pi+pi0WZ}
\eea
and\l
$\bullet$
\begin{equation}
{\cal S}_{K_s \rar \pi^+\pi^-} = -4i s_\theta c_\theta f M_W^2 \sqrt{2}G_F A
\left((J1-J2)(M_W,m_H,m_P)+(J1-J2)(M_W,m_S,m_P)\right);
\label{eq:SKs1}
\end{equation}
$\bullet$
\begin{equation}
{\cal S}_{K_s \rar \pi^0\pi^0} = -4i s_\theta c_\theta f M_W^2 \sqrt{2}G_F A
\left(2(J1-J2)(M_W,m_S,m_P)+(J1-J2)(M_W,m_S,m_G)\right).
\label{eq:SKs2}
\end{equation}
$A$ is the dimensionless coupling constant of the proposed model of
interactions
\begin{equation}
A= \Lambda \sqrt{\sqrt{2}G_F};
\label{eq:Aint}
\end{equation}
its electroweak counterpart is
\begin{equation}
A_{EW} = M_W \sqrt{\sqrt{2}G_F} \approx 0.324.
\label{eq:Aew}
\end{equation}
The two functions $J1(M,m_1,m_2)$ and $J2(M,m_1,m_2)$ are defined by
\bea
J1(M,m_1,m_2)&=&p.p_1 \int\frac{d^4q}{(2\pi)^4}
    \frac{1}{(q-p_1)^2 - m_1^2)}\frac{1}{(q-p)^2-m^2)}\frac{1}{q^2-M^2},\cr
J2(M,m_1,m_2)&=&\int\frac{d^4q}{(2\pi)^4}(q.p)(q.p_1)
        \frac{1}{(q-p_1)^2-m_1^2)}\frac{1}{(q-p)^2-m^2)}
        \frac{1}{q^2-M^2}\frac{1}{q^2};
\label{eq:J1J2}
\eea
we always work in the limit $m_\pi, m_K \ll m_1$ or $m_2$.

One then has the analytic expressions:
\begin{equation}
J1(M,m_1,m_2)=\frac{i}{16\pi^2}\frac{1}{2}\frac{m_K^2}{m_1^2-m_2^2}
  \left(\frac{m_2^2\ln{m_2^2}-M^2\ln{M^2}}{m_2^2-M^2}
       -\frac{m_1^2\ln{m_1^2}-M^2\ln{M^2}}{m_1^2-M^2} \right),
\label{eq:J1}
\end{equation}
\begin{equation}
J2(M,m_1,m_2)=\frac{1}{2}J1(M,m_1,m_2)
+\frac{i}{16\pi^2}\frac{1}{8}\frac{m_K^2}{m_1^2-m_2^2}
  \left(\ln{\frac{m_1^2}{m_2^2}}+\frac{M^2}{M^2-m_1^2}\ln{\frac{M^2}{m_1^2}}
   -\frac{M^2}{M^2-m_2^2}\ln{\frac{M^2}{m_2^2}}\right),
\label{eq:J2}
\end{equation}
leading to the relevant combination (using $p.p_1=(1/2)m_K^2$)
\begin{equation}
(J1-J2)(M,m_1,m_2)=\frac{i}{16\pi^2}\frac{3}{8}\frac{m_K^2}{m_1^2-m_2^2}
\left(\frac{m_1^2\ln{m_1^2}}{M^2-m_1^2}-\frac{m_2^2\ln{m_2^2}}{M^2-m_2^2}
 -\frac{(m_1^2-m_2^2)M^2\ln{M^2}}{(M^2-m_1^2)(M^2-m_2^2)}\right).
\label{eq:J1-J2}
\end{equation}
The two following limits will be used:
\bea
(J1-J2)(M,m_1,0)&=&\frac{i}{16\pi^2}\frac{3}{8}\frac{m_K^2}{M^2-m_1^2}
           \ln{\frac{m_1^2}{M^2}},\cr
(J1-J2)(M,m,m) &=& \frac{i}{16\pi^2}\frac{3}{8}\frac{m_K^2}{M^2-m^2}
  \left(1+\frac{M^2}{M^2-m^2}\ln{\frac{m^2}{M^2}}\right).
\label{eq:Jlimits}
\eea
Two facts are noteworthy at this stage:\l
- $K_s \rar \pi^0\pi^0$ decays are restored;\l
- while only the $W$ plays a role in $K_s \rar \pi\pi$ decays, $W$ and $Z$
mediate $K^+ \rar \pi^+\pi^0$ transitions, with {\em a priori} competing
signs.

\subsubsection{The limit of flavour and parity symmetry for intermediate
states}
\label{subsub:symmetric}

Let $m$ the unique mass of all intermediate electroweak states
(supposed to be higher than the pion and kaon masses).
Then, eqs.~(\ref{eq:SK+pi+pi0},\ref{eq:SK+pi+pi0WZ},
\ref{eq:SKs1},{\ref{eq:SKs2}) reduce to
\bea
{\cal S}_{K^+ \rar \pi^+\pi^0} &=&-8i s_\theta c_\theta f M_W^2 \sqrt{2}G_F A
\left((J1-J2)(M_W,m,m)-\frac{1}{c_W^2}(J1-J2)(M_Z,m,m)\right);\cr
{\cal S}_{K_s \rar \pi^+\pi^-}&=& -8i s_\theta c_\theta f M_W^2 \sqrt{2}G_F A
                          (J1-J2)(M_W,m,m);\cr
{\cal S}_{K_s \rar \pi^0\pi^0}&=&\frac{3}{2}{\cal S}_{K_s \rar \pi^+\pi^-},
\label{eq:Ssym}
\eea
where $(J1-J2)(M_{W,Z},m,m)$ is given by eq.~(\ref{eq:Jlimits}).

The mechanism which damps ${\cal S}_{K^+ \rar \pi^+\pi^0}$ with respect to
the two decay amplitudes of the neutral kaons is conspicuous: the $Z$ gauge
boson only plays a role in the former and comes with a relative negative
sign with respect to the $W$ contribution. This cancelation is complete for
$\cos\theta_W=1$ equivalent to $M_W = M_Z$.

On fig.~5 below the ratio of the two amplitudes ${\cal S}_{K_s \rar
\pi^+\pi^-}/{\cal S}_{K^+ \rar \pi^+\pi^0}$ is plotted as a function of $m$:
this ratio is independent of the coupling constant $A$.
All dependence on the mass of the external legs also divides out.

To take consistently this symmetric limit requires in particular that it be
also taken for external legs, and thus that $m_K = m_\pi$: the
pure electroweak contribution at the tree level vanishes at this limit.

\vbox{
\figskip
\bct
\epsfig{file=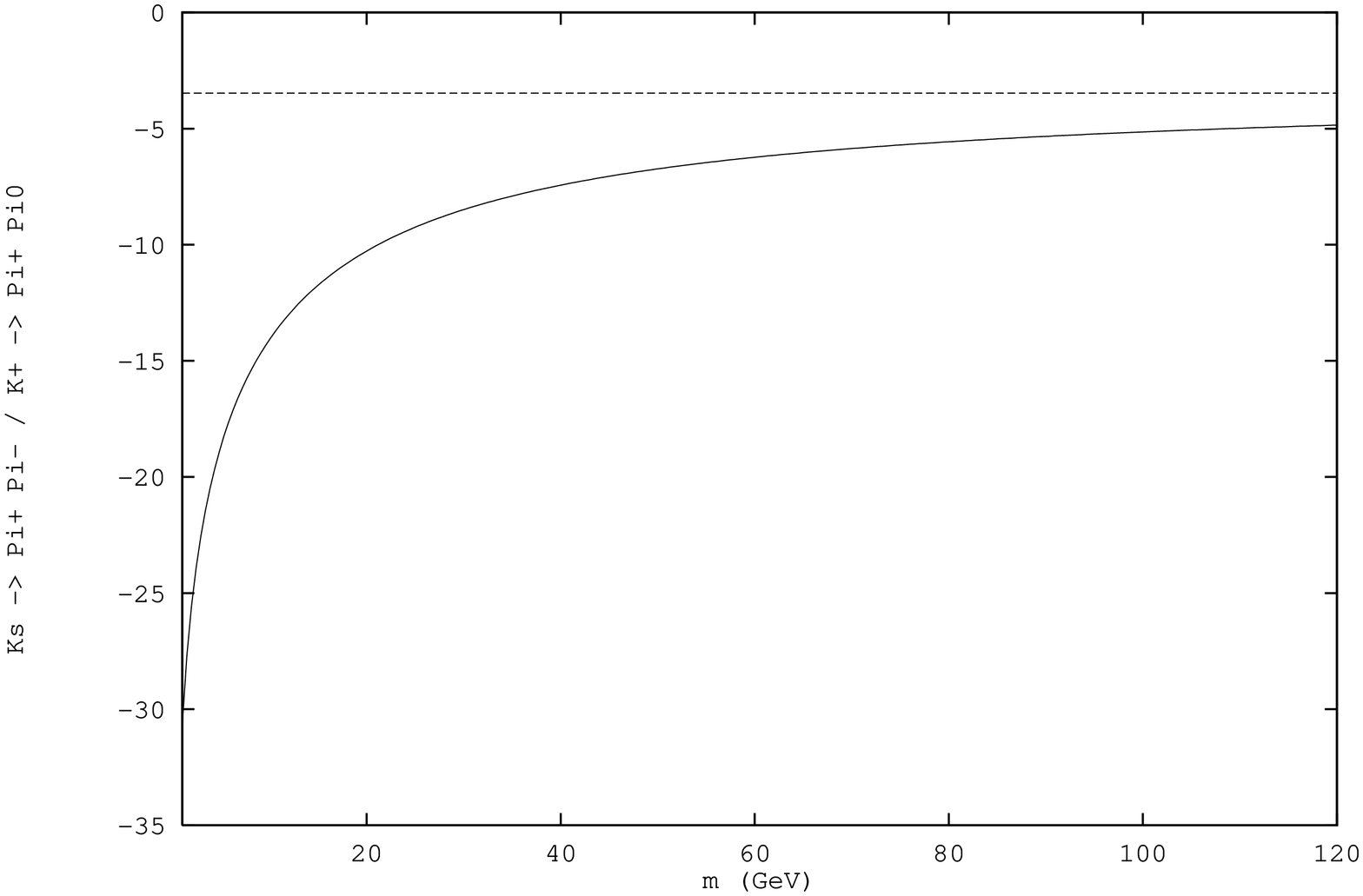,height=15truecm,width=8truecm,angle=-90}
\figskip
{\em Fig.~5: ratio of amplitudes ${\cal S}_{K_s \rar \pi^+\pi^-}
/{\cal S}_{K^+ \rar \pi^+\pi^0}$ at the limit of degenerate electroweak
eigenstates.}
\ect
}
At the limit $m \rar \infty$ the ratio ${\cal S}_{K_s \rar \pi^+\pi^-}/
{\cal S}_{K^+ \rar\pi^+\pi^0}$ goes to
\begin{equation}
\frac{{\cal S}_{K_s \rar \pi^+\pi^-}}{{\cal S}_{K^+ \rar\pi^+\pi^0}}
\underset{m\rar\infty}{\longrightarrow}
1 - \frac{(J1-J2)(M_Z,m,m)}{(J1-J2)(M_W,m,m)}= -\frac{1}{\tan^2\theta_W}.
\label{eq:Rsym}
\end{equation}
%
\subsubsection{Splitting scalar and pseudoscalar intermediate states}
\label{subsub:2scales}

The next, less drastic approximation, is to allow a splitting between scalar
and pseudoscalar states, and thus to introduce two mass scales for intermediate
states, $m_P$ and $m_S$.
We shall still consider $m_\pi, m_K \ll m_P$ or  $m_S$.

The consistency of this limit again requires that one take $m_K = m_\pi$,
making the pure electroweak amplitudes vanish.

The expressions for the $K\rar \pi\pi$ amplitudes become:
\bea
{\cal S}_{K^+ \rar \pi^+\pi^0} &=&-8i s_\theta c_\theta f M_W^2 \sqrt{2}G_F A
\left((J1-J2)(M_W,m_S,m_P)-\frac{1}{c_W^2}(J1-J2)(M_Z,m_S,m_P)\right);\cr
{\cal S}_{K_s \rar \pi^+\pi^-}&=& -8i s_\theta c_\theta f M_W^2 \sqrt{2}G_F A
                          (J1-J2)(M_W,m_S,m_P);\cr
{\cal S}_{K_s \rar \pi^0\pi^0}&=&\frac{3}{2}{\cal S}_{K_s \rar \pi^+\pi^-},
\label{eq:S2scales}
\eea
and the ratio ${{\cal S}_{K_s \rar \pi^+\pi^-}}/{{\cal S}_{K^+
\rar\pi^+\pi^0}}$ is plotted on fig.~6 below as  function of $m_P$ for
and $m_S$; we have made $m_P$ vary in the interval $[1\,GeV,10\,GeV]$ and $m_S$
vary in the interval $[1\,GeV, 200\,GeV]$.

\vbox{
\figskip
\bct
\epsfig{file=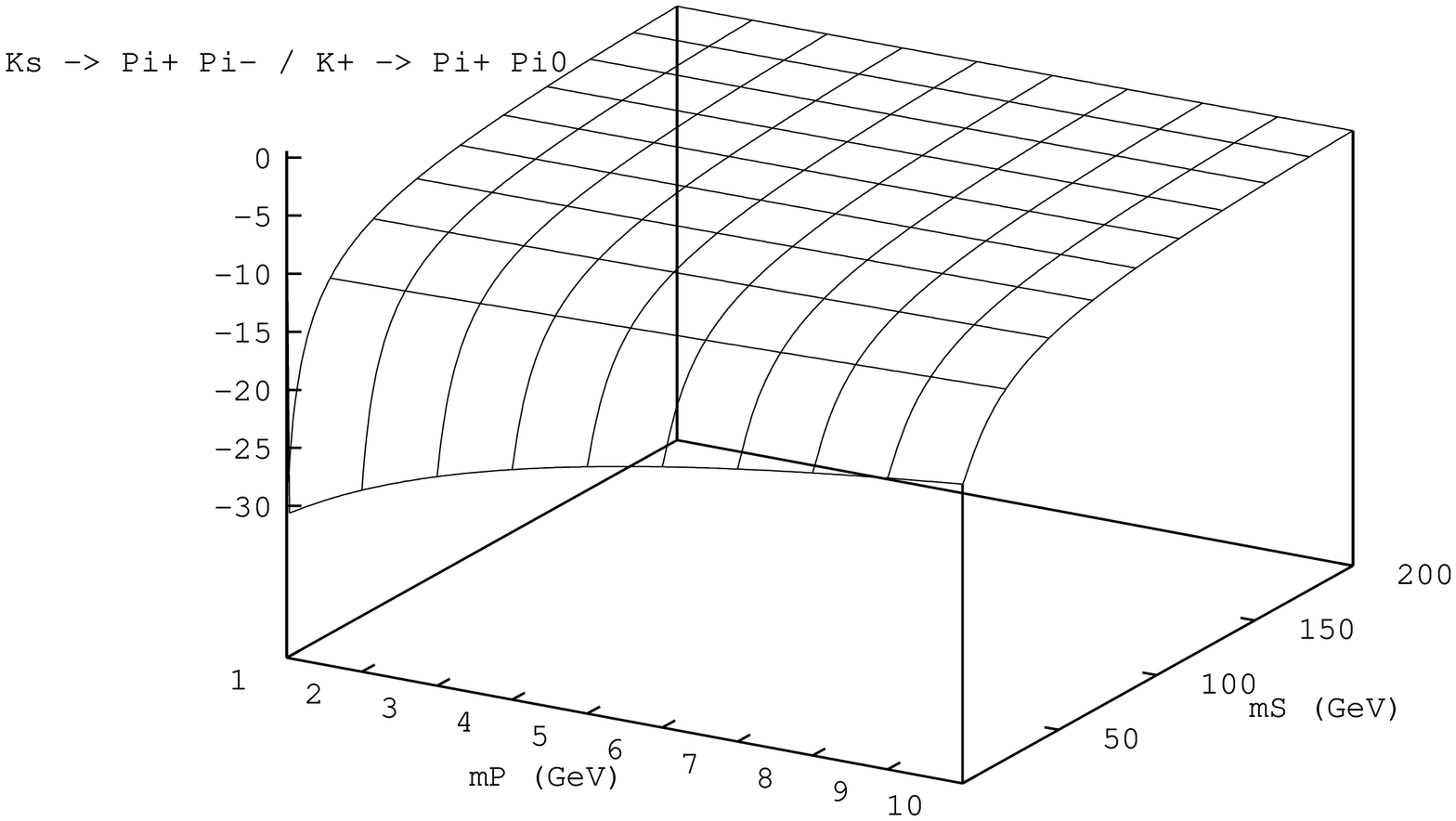,height=15truecm,width=8truecm,angle=-90}
\figskip
{\em Fig.~6: ratio of amplitudes
${\cal S}_{K_s \rar \pi^+\pi^-}
/{\cal S}_{K^+ \rar \pi^+\pi^0}$ as a function of $m_P$ and $m_S$.}
\ect
}
When $m_S \rar \infty$, the above ratio has the same asymptotic value
$-1/\tan^2\theta_W$ as in the previous case.

\subsubsection{A more realistic approximation}
\label{subsub:4scales}

We now work with the four scales $m_G\equiv 0, m_P, m_S, m_H$.

Of course, this approximation is still very crude as it does not take into
account the mass splittings between electroweak eigenstates of the same
parity.

\paragraph{The ratio of $\boldsymbol{K_s \rar \pi^+\pi^-}$ over
$\boldsymbol{K^+ \rar \pi^+\pi^0}$ amplitudes.}
\label{parag;Kspi+pi-/K+pi+pi0}

The $W-Z$ cancelation still operates, and now the amplitude for $K^+ \rar
\pi^+\pi^0$ can even vanish. Its zeroes determine the maximum violation of
strong isospin. The equation which determines them  can unfortunately
only be solved numerically.

On fig.~7 below is plotted the
ratio of amplitudes ${\cal S}_{K_s \rar \pi^+\pi^-}
/{\cal S}_{K^+ \rar \pi^+\pi^0}$ for $m_P = 1\,GeV$, 3 values of
$m_S$, $m_S = 1.5\,GeV, 10\,GeV$ and $50\,GeV$, and for $m_H \in [0,200\,GeV]$.

For given values of $(m_P, m_S)$, the above ratio is seen to be a sensitive
function of the Higgs mass.

\vbox{
\figskip
\bct
\epsfig{file=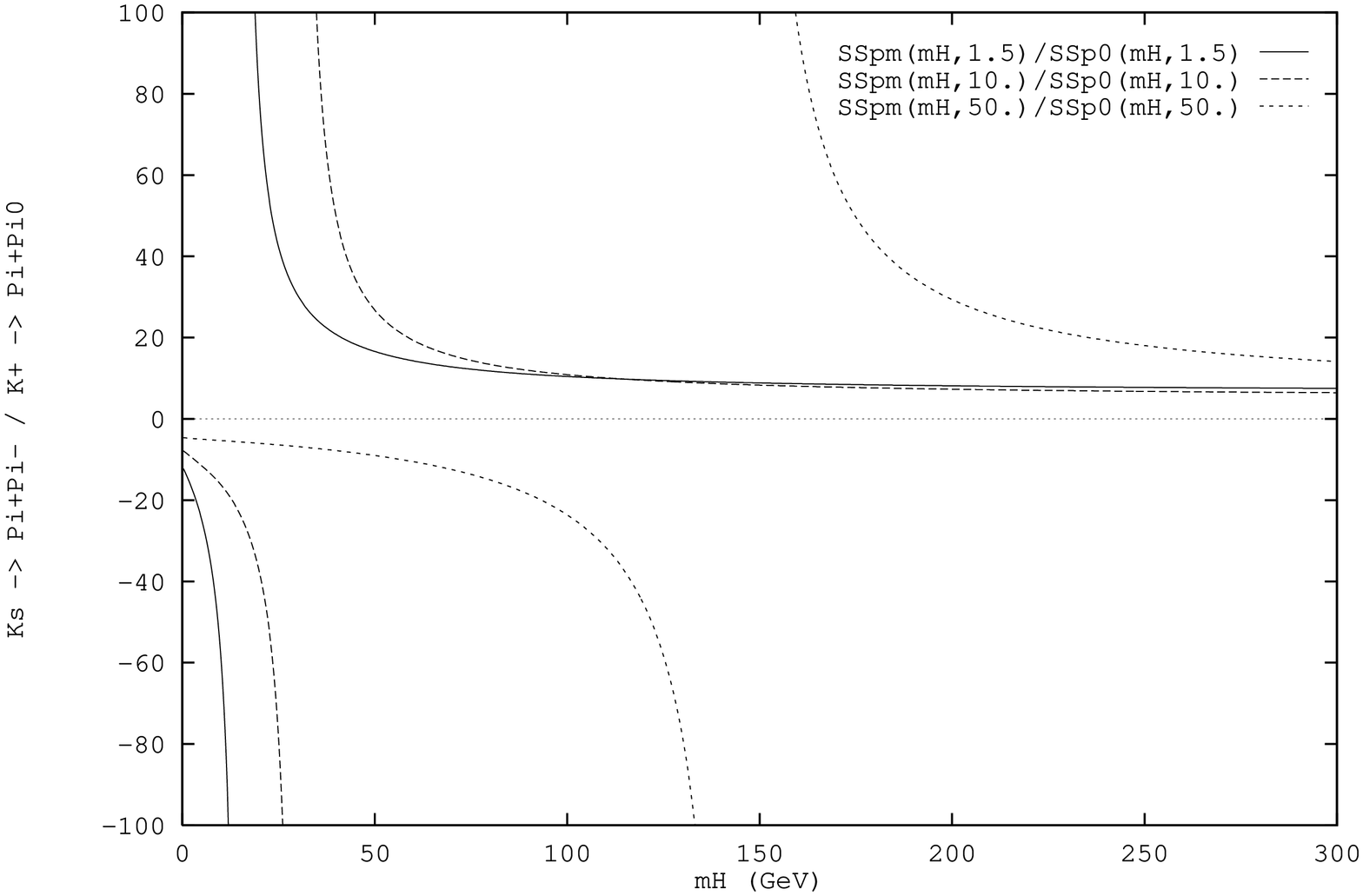,height=15truecm,width=8truecm,angle=-90}
\figskip
{\em Fig.~7: ratio of amplitudes
${\cal S}_{K_s \rar \pi^+\pi^-}
/{\cal S}_{K^+ \rar \pi^+\pi^0}$ for $m_P= 1\,GeV$, $m_S=1.5, 10$ and
$50\,GeV$.}
\ect
}
%
\paragraph{The ratio of $\boldsymbol{K_s \rar \pi^+\pi^-}$ over
$\boldsymbol{K_s \rar \pi^0\pi^0}$ amplitudes.}
\label{parag:Kspi+pi-/Kspi0pi0}

It is no longer fixed as it was in the two previous approximations.
On fig.~8 below, it is plotted as a function of $m_S \in [1.5\, GeV,
50\,GeV]$ and $m_H \in [5\, GeV, 200\,GeV]$ for $m_P=1\,GeV$.

\vbox{
\figskip
\bct
\epsfig{file=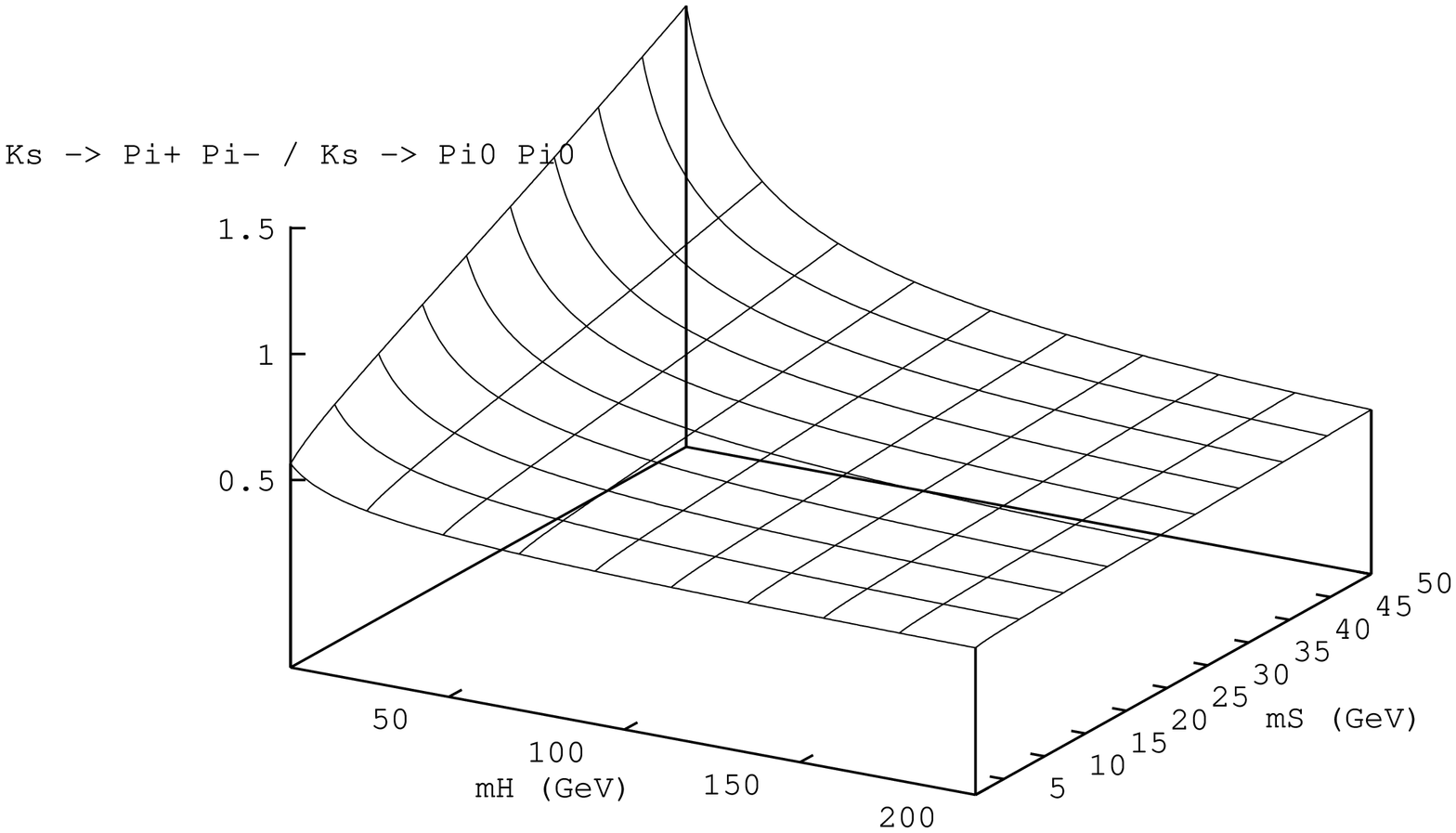,height=15truecm,width=8truecm,angle=-90}
\figskip
{\em Fig.~8: ratio of amplitudes
${\cal S}_{K_s \rar \pi^+\pi^-}
/{\cal S}_{K_s \rar \pi^0\pi^0}$ for $m_P= 1 GeV$.}
\ect
}
The experimental value $\sqrt{2}$ for this ratio, which is another aspect of
the $\Delta I = 1/2$ rule, is seen to be easily reproducible, too.

\paragraph{A short comment on the mass splittings.}

All the numerical computations that we presented seem to favour large mass
splittings between the multiplets and still larger between scalar and
pseudoscalar mesons. We have taken advantage of the arbitrariness of the
different electroweak mass scales which is allowed by the model, and of the
supplementary splitting between scalars and pseudoscalars which is triggered
by the breaking of the electroweak and chiral symmetries, while preserving
the custodial $SU(2)_V$ symmetry. Experimentally, the scalar
mesons are indeed observed in general to have masses much higher than the
pseudoscalars.
\paragraph{$\boldsymbol{\Lambda}$ or ``strong'' interactions?}
\label{parag:strong?}

The goal of this work is not to describe precisely $K \rar \pi\pi$ decays
individually, but to uncover the mechanism at the origin of large isospin
breaking in dimensionless ratios.
And indeed, the number of parameters is too large (16 mass scales and one
coupling constant) and there are too many unknown to go with reasonable
confidence beyond the general approximations and
results that have been exposed above.

However, let us get some idea of what would be, within the framework
of the last approximation, the value of $A$ that could reproduce the order of
magnitude of the modulus  of the amplitude for $K_s \rar \pi^+\pi^-$, the
experimental value of which is
\begin{equation}
{\cal A}^{exp}_{K_s \rar \pi^+\pi^-} \approx 3.92\ 10^{-7}GeV.
\end{equation}
Our prediction is plotted on fig.~9 as a function of $A$ and $m_H$,
for $m_P = 1\,GeV$ and $m_S = 10\,GeV$.

\vbox{
\figskip
\bct
\epsfig{file=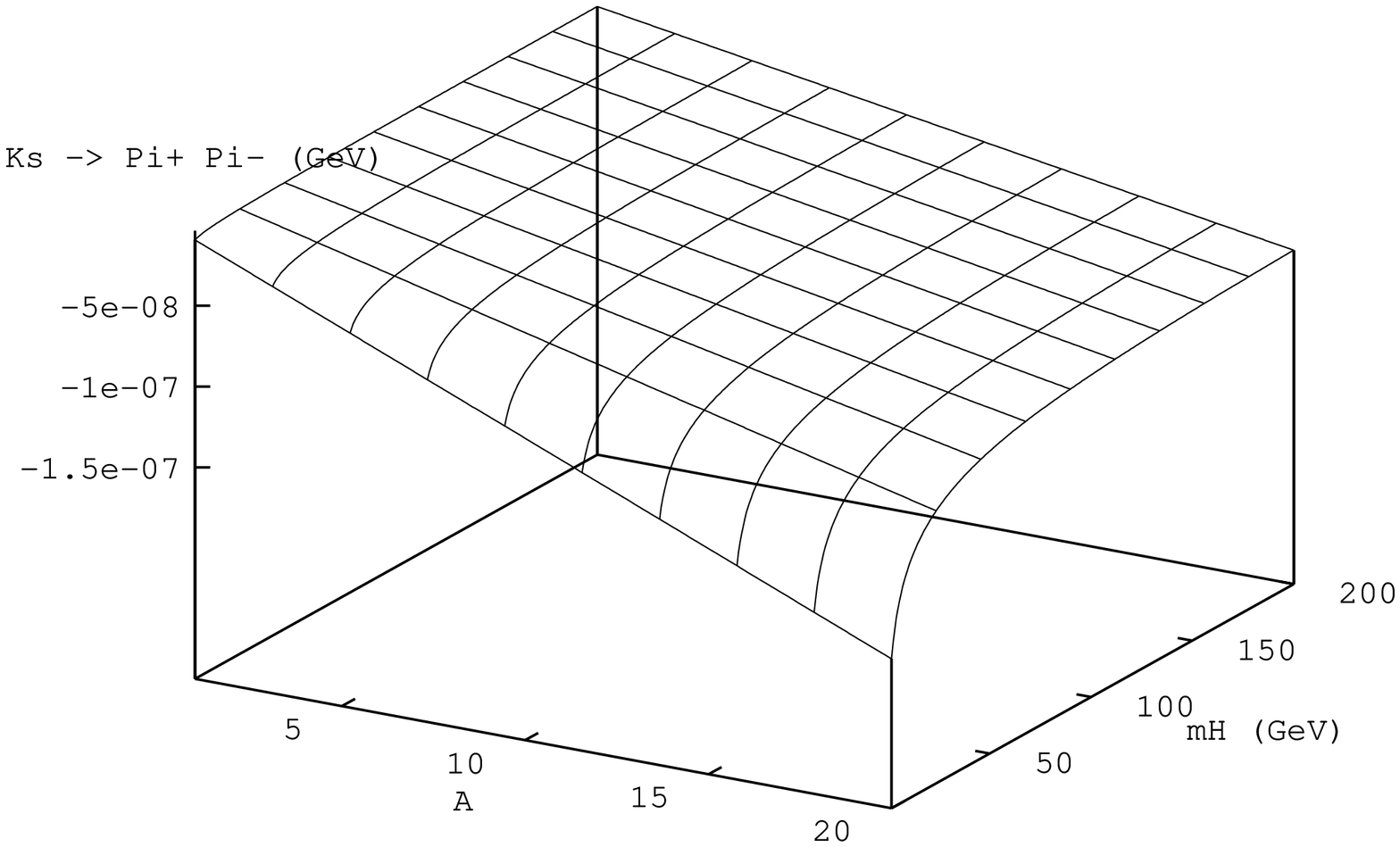,height=15truecm,width=8truecm,angle=-90}
\figskip
{\em Fig.~9: amplitude ${\cal S}_{K_s \rar \pi^+\pi^-}(GeV)$ for $m_P =
1\ GeV$ and $m_S = 10\,GeV$.}
\ect
}
The only conclusion that we can draw is that large values of $A$ seem to be
favored, which correspond to ``strong'' interactions.

\section{Conclusion}
\label{section:conclusion}

A clear and simple mechanism yielding large violation of the isospin
symmetry in $K \rar \pi\pi$ decays has been uncovered.

Results have been obtained in a model which respects all known symmetries
of the interacting mesons  and  which describes them as both the fields
and the particles of the corresponding Lagrangian.
Quarks have only been considered as mathematical objects \cite{GellMann},
which circumvents in particular the problems of confinement and
hadronization.

That it stays operative when the proposed $\Lambda$-interactions become
strong cannot of course be assessed since we only performed  one-loop
computations, and only relies on the fact that the one-loop ratios
that we have been considering do not depend on $\Lambda$.
But such a problem already existed in the QCD approach, and the mechanism for
isospin violation was by far less conspicuous there.

We have seen that the only study of $K \rar \pi\pi$ decays is not enough to
give trustworthy predictions concerning the mass of the Higgs boson, though it
stays among the main goals to be achieved. It is conceivable that similar
studies for other nonleptonic decays might eliminate some unknown
parameters that we encountered here (like the spectrum of the scalar mesons)
and provide more than just qualitative results.

A natural application of the amplification by ``strong'' interactions
of an isospin violation due to electroweak interactions
concerns the always problematic isospin violating decays of the
$\eta$ meson into three pions \cite{eta3pi}. By comparison with the decays
studied above, two major sources of complications are expected to make the
computations much more tedious: the final state with three pions which can
be generated in many ways and thus can arise from more numerous Feynman
diagrams, and the uncertainty concerning the flavour structure of the
incoming state and its mixing with other neutral flavour diagonal
pseudoscalar mesons.

Another natural extension concerns the case of three generations and the
mechanism for $CP$ violation. The length of the computations as performed in
the framework above would make their automation compulsory,
and we did not find yet the appropriate method. Furthermore, the most
general electroweak mass eigenstates can be linear combinations of the
quadruplets (\ref{eq:SP},\ref{eq:PS}) displayed in section
\ref{section:theory}, which introduces, in
addition to the CKM matrix controlling the embedding of the electroweak group
into the chiral group, another mixing matrix at the mesonic level; it was
also shown in \cite{Machet3} that the presence of a complex CKM matrix is no
longer a sufficient condition to have indirect $CP$ violation, and that this
phenomenon can now happen, unlike in the Glashow-Salam-Weinberg model, with
two generations only through the new mixing matrix just mentioned.
One thus has to face the existence of two {\it a priori} unconnected mixing
matrices and the large spectrum of possibilities which arise
makes the detailed investigation of $CP$ violation a very demanding task.

These points are currently under investigation; I just hope to have given hints
in the present work that the story of $K \rar \pi\pi$ decays still proceeds
and that the exposed technique can provide valuable information
on hadronic interactions.
\vskip 1cm
{\it \underline {Acknowledgments}: I would like to thank the referee of
this work for his careful study of the paper, his open-mindedness
and his constructive questions and comments.}
%
\newpage\null
\listoffigures
\bigskip
\begin{em}
Fig.~1: $K^+ \rar \pi^+ \pi^0$ and $K_s \rar \pi^+ \pi^-$ electroweak
decays at tree level;\l
Fig.~2: non-vanishing electroweak 1-loop diagrams for $K^+ \rar \pi^+
\pi^0$;\l
Fig.~3: the ratio of one-loop over tree amplitudes for $K^+ \rar
\pi^+\pi^0$ in the pure electroweak case;\l
Fig.~4: trilinear meson coupling induced by $\Lambda$-interactions;\l
Fig.~5: ratio of amplitudes ${\cal S}_{K_s \rar \pi^+\pi^-}
             /{\cal S}_{K^+ \rar \pi^+\pi^0}$ at the limit of degenerate
              electroweak eigenstates;\l
Fig.~6: ratio of amplitudes
    ${\cal S}_{K_s \rar \pi^+\pi^-} /{\cal S}_{K^+ \rar \pi^+\pi^0}$
    as a function of $m_P$ and $m_S$;\l
Fig.~7: ratio of amplitudes
    ${\cal S}_{K_s \rar \pi^+\pi^-} /{\cal S}_{K^+ \rar \pi^+\pi^0}$
    with four mass scales;\l
Fig.~8: ratio of amplitudes
    ${\cal S}_{K_s \rar \pi^+\pi^-} /{\cal S}_{K_s \rar \pi^0\pi^0}$
    with four mass scales;\l
Fig.~9: amplitude ${\cal S}_{K_s \rar \pi^+\pi^-}$;\l
Fig.~10: vanishing 1-loop electroweak diagrams for $K^+ \rar \pi^+ \pi^0$;\l
Fig.~11: non-vanishing diagrams for $K^+ \rar \pi^+\pi^0$
     including a trilinear $\Lambda$ coupling;\l
Fig.~12: vanishing or irrelevant diagrams for $K^+ \rar \pi^+\pi^0$
     including a trilinear $\Lambda$ coupling;\l
Fig.~13: non-vanishing diagrams for $K_s \rar \pi\pi$
     including a trilinear $\Lambda$ coupling.
\end{em}
\newpage\null
{\Large\bf Appendix}
%
\appendix
\section{Representations}
\label{app:reps}
The four types of stable quadruplets are $\Phi({\mathbb D}_1)$ given by
eq.~(\ref{eq:PHI1}) and $\Phi({\mathbb D}_2)$, $\Phi({\mathbb D}_3)$ and
$\Phi({\mathbb D}_4)$ given below.

\vbox{
\bea
& &\Phi({\mathbb D}_2) = \cr
& & \hskip -2cm \left[
\frac{1}{\sqrt{2}} \left(\ba{rrcrr}
     1 &   &\vline &    &    \nonumber\\
       & -1 &\vline &    &    \nonumber\\
     \hline
  &  & \vline & c_\theta^2 - s_\theta^2 & 2c_\theta s_\theta  \nonumber\\
  &  & \vline & 2c_\theta s_\theta & s_\theta^2 -c_\theta^2  \ea \right),
\frac{i}{\sqrt{2}} \left(\ba{rrcrr}
             1 &    & \vline &   &   \nonumber\\
               & -1 & \vline &   &   \nonumber\\
             \hline
      &  & \vline & s_\theta^2 - c_\theta^2 & -2c_\theta s_\theta
\nonumber\\
      &  & \vline & -2c_\theta s_\theta & c_\theta^2 - s_\theta^2  \ea
\right),
\right .\nonumber\\
& & \hskip 7cm \left .
i \left(\ba{rrcrr}   &  &\vline & c_\theta & s_\theta \nonumber\\
                             &  &\vline & s_\theta & -c_\theta \nonumber\\
                            \hline
                             &  &\vline &   &     \nonumber\\
                             &  &\vline &   &  \ea \right),
i \left(\ba{rrcrr}     &   &\vline &   &   \nonumber\\
                               &   &\vline &   &   \nonumber\\
                              \hline
                              c_\theta & s_\theta &\vline &   & \nonumber\\
                              s_\theta & -c_\theta &\vline &   &   \ea
\right)
\right]; \nonumber\\
& &
\label{eq:PHI2}
\eea
}
\vbox{
\bea
& &\Phi({\mathbb D}_3) = \cr
& & \hskip -2cm \left[
\frac{1}{\sqrt{2}} \left(\ba{rrcrr}
        & 1 &\vline &    &    \nonumber\\
      1 &  &\vline &    &    \nonumber\\
     \hline
     &  & \vline & -2c_\theta s_\theta & c_\theta^2 - s_\theta^2
\nonumber\\
     &  & \vline &  c_\theta^2 - s_\theta^2 & 2c_\theta s_\theta   \ea
\right),
\frac{i}{\sqrt{2}} \left(\ba{rrcrr}
         & 1 & \vline &   &   \nonumber\\
       1 &  & \vline &   &   \nonumber\\
       \hline
     &  & \vline & 2c_\theta s_\theta & s_\theta^2 - c_\theta^2 \nonumber\\
     &  & \vline & s_\theta^2 - c_\theta^2 & -2c_\theta s_\theta \ea
\right),
\right .\nonumber\\
& & \hskip 7cm \left .
i \left(\ba{rrcrr}   &  & \vline & -s_\theta & c_\theta \nonumber\\
                             &  & \vline & c_\theta &  s_\theta \nonumber\\
                            \hline
                            &  &\vline &   &     \nonumber\\
                            &  &\vline &   &  \ea \right),
i \left(\ba{rrcrr}     &   &\vline &   &   \nonumber\\
                               &   &\vline &   &   \nonumber\\
                              \hline
                             -s_\theta & c_\theta &\vline &   &
\nonumber\\
                              c_\theta & s_\theta &\vline &   &   \ea
\right)
\right]; \nonumber\\
& &
\label{eq:PHI3}
\eea
}
\vbox{
\bea
& &\Phi({\mathbb D}_4) = \cr
& & \hskip -2cm \left[
\frac{1}{\sqrt{2}} \left(\ba{rrcrr}           & 1 &\vline &    & \nonumber\\
                                         -1 &   &\vline &    & \nonumber\\
                                            \hline
                                            &   &\vline &    & 1
\nonumber\\
                                            &   &\vline & -1 &    \ea
\right),
\frac{i}{\sqrt{2}}\left(\ba{rrcrr}   & 1 &\vline &   &   \nonumber\\
                                         -1 &   &\vline &   &   \nonumber\\
                                           \hline
                                            &   &\vline &   & -1
\nonumber\\
                                            &   &\vline & 1 &    \ea
\right),
 i \left(\ba{rrcrr}   &  &\vline & -s_\theta & c_\theta \nonumber\\
                            &  &\vline & -c_\theta & -s_\theta \nonumber\\
                            \hline
                            &  &\vline &   &     \nonumber\\
                            &  &\vline &   &  \ea \right),
i  \left(\ba{rrcrr}     &   &\vline &   &   \nonumber\\
                              &   &\vline &   &   \nonumber\\
                            \hline
                           s_\theta & c_\theta &\vline &   &   \nonumber\\
                           -c_\theta & s_\theta &\vline &   &   \ea \right)
\right]. \nonumber\\
& &
\label{eq:PHI4}
\eea
}

%
\section{Normalizing the fields and the Lagrangian}
\label{app:normalizing}

The Lagrangian is:
\begin{equation}
{\cal L}= {\cal L}_{gauge} + {\cal L}(\Phi) +{\cal L}_{fermions}.
\label{eq:Lag}
\end{equation}
The gauge field Lagrangian in eq.~(\ref{eq:Lag}) is
\begin{equation} 
{\cal L}_{gauge} = \sum_{gauge fields} -\frac{1}{4} F^{\mu\nu}F_{\mu\nu},
\label{eq:Lgauge}
\end{equation}
and
${\cal L}_{fermions}$ is the usual Glashow-Salam-Weinberg Lagrangian for
fermions (leptons) \cite{GlashowSalamWeinberg}.

The Lagrangian for scalars ${\cal L}(\Phi)$ is chosen to be
(note the signs for the various kinetic terms)
\bea
{\cal L}(\Phi) &=& \frac{1}{4} D_\mu \varphi({\mathbb D}_1) \otimes
                            D^\mu \varphi({\mathbb D}_1)
                    -V_1(\varphi({\mathbb D}_1))
              + \frac{1}{4} D_\mu \varphi({\mathbb D}_2) \otimes
                            D^\mu \varphi({\mathbb D}_2)
                  -  V_2(\varphi({\mathbb D}_2)) \cr
&& \cr
              &+& \frac{1}{4} D_\mu \varphi({\mathbb D}_3) \otimes
                            D^\mu \varphi({\mathbb D}_3)
                  -  V_3(\varphi({\mathbb D}_3))
              - \frac{1}{4} D_\mu \varphi({\mathbb D}_4) \otimes
                            D^\mu \varphi({\mathbb D}_4)
                  -  V_4(\varphi({\mathbb D}_4)) \cr
&& \cr
              &-& \frac{1}{4} D_\mu \chi({\mathbb D}_1) \otimes
                            D^\mu \chi({\mathbb D}_1)
                  -  U_1(\chi({\mathbb D}_1))
              - \frac{1}{4} D_\mu \chi({\mathbb D}_2) \otimes
                            D^\mu \chi({\mathbb D}_2)
                  -  U_2(\chi({\mathbb D}_2)) \cr
&& \cr
              &-& \frac{1}{4} D_\mu \chi({\mathbb D}_3) \otimes
                            D^\mu \chi({\mathbb D}_3)
                  -  U_3(\chi({\mathbb D}_3))
              + \frac{1}{4} D_\mu \chi({\mathbb D}_4) \otimes
                            D^\mu \chi({\mathbb D}_4)
                  -  U_4(\chi({\mathbb D}_4)).\cr
&&
\label{eq:Lphi}
\eea
The choice (\ref{eq:Dmatrix}) of $\mathbb D$ matrices makes the kinetic terms of
${\cal L}_\Phi$ diagonal both in the electroweak eigenstates and in the
flavour or ``strong'' eigenstates (corresponding to $\mathbb M$ matrices
with only one non-vanishing entry equal to $1$).

The normalization $1/4$ for the $\Phi$ fields in eq.~(\ref{eq:Lphi})
yields the usual $1/2$ when
the Lagrangian is written for the flavour eigenstates
(see eqs.~(\ref{eq:kinetic}) and (\ref{eq:scaleL})).

This entails however that the propagators of the $\mathbb S$ and $\mathbb P$
electroweak eigenstates get a
factor $2$ in their numerators; for example, the propagator of the neutral
goldstone ${\mathbb P}^3({\mathbb D}_1)$ is
\begin{equation}
D_{{\mathbb P}^3} (q) = \frac {2i}{q^2}.
\label{eq:DP}
\end{equation}
$V_2,V_3,V_4,U_1,U_2,U_3,U_4$ are quartic potentials which do not trigger
symmetry breaking, while
\begin{equation}
V_1(\varphi({\mathbb D}_1) = -\frac{\sigma^2}{4}\varphi({\mathbb D}_1) \otimes
                          \varphi({\mathbb D}_1)
     + \frac{\lambda}{8} \left(\varphi({\mathbb D}_1) \otimes
                   \varphi({\mathbb D}_1)\right)^{\otimes 2}
\label{eq:V1}
\end{equation}
does.

The Higgs boson is shifted according to
\begin{equation}
H = \la H \ra + h,\ \la H \ra = \frac{v}{\sqrt{2}}.
\label{eq:Higgs}
\end{equation}
Using eq.~(34) of \cite{Machet1} for the covariant derivatives of the fields,
one gets
\begin{equation}
M_W^2 = \frac{g^2\,v^2}{16}.
\label{eq:MW}
\end{equation}
The low energy relation coming from $e^+\,e^-$ scattering
\begin{equation}
\frac{G_F}{\sqrt{2}} = \frac{g^2}{8M_W^2}
\label{eq:e+e-}
\end{equation}
yields, together with eq.~(\ref{eq:MW}),
\begin{equation}
G_F\,v^2 = 2\sqrt{2}.
\label{eq:GF}
\end{equation}
To make the link with observed mesons, the fields $\mathbb M$ have to be
normalized with the factor $a$ of eq.~(\ref{eq:a}):
\begin{equation}
{\mathbb M} = a \tilde{\mathbb M},
\label{eq:scaleM}
\end{equation}
such that, for example
\begin{equation}
\tilde{\mathbb P}^+({\mathbb D}_1)
= \frac{1}{a}\  {\mathbb P}^+({\mathbb D}_1)
               =c_\theta(\pi^+ + D_s^+) + s_\theta(K^+ - D^+).
\label{eq:P1}
\end{equation}
The gauge fields are likewise rescaled according to
\bea
W_\mu &=& a \tilde W_\mu, \cr
Z_\mu &=& a \tilde Z_\mu.
\label{eq:scaleW}
\eea
The Lagrangian is then globally rescaled by $1/a^2$ such that the
kinetic terms for the observed mesons are normalized to the usual $1/2$
factor
\bea
\tilde{\cal L}(\tilde\Phi) \equiv\frac{{\cal L}(a\tilde\Phi)}{a^2}
&=& \frac{1}{4} D_\mu \tilde\varphi({\mathbb D}_1) \otimes
                  D^\mu \tilde\varphi({\mathbb D}_1) +\cdots
                    -\tilde V_1(\tilde\varphi({\mathbb D}_1)) -\cdots.\cr
&& \cr
&=& \frac{1}{2}(D_\mu\pi^0 D^\mu\pi^0 + 2\,D_\mu \pi^+ D^\mu \pi^-
         + D_\mu K^0 D^\mu \ol{K^0} + 2\,D_\mu  K^+ D^\mu  K^- +\cdots)\cr
&& -\tilde V_1(\tilde\varphi({\mathbb D}_1)) -\cdots
\label{eq:scaleL}
\eea
The quartic potential $V_1(\varphi({\mathbb D}_1))$ has become
\begin {equation}
\tilde V_1 (\tilde\varphi({\mathbb D}_1))
 = -\frac{\sigma^2}{4}\tilde\varphi({\mathbb D}_1)^{\otimes 2}
+ a^2 \frac{\lambda}{8} \tilde\varphi({\mathbb D}_1)^{\otimes 4},
\label{eq:scaleV}
\end{equation}
and generates in particular the vertex (including the ``$i$'' of
$i\, (-\tilde V)$ occurring in $e^{iS}$)
\begin{equation}
-ia \frac{\lambda v}{2\sqrt{2}} \tilde h\otimes
\tilde{\vec{\mathbb P}}({\mathbb D}_1)\otimes
\tilde{\vec{\mathbb P}}({\mathbb D}_1).
\label{eq:vertex}
\end{equation}
$\tilde h$ is defined by
\begin{equation}
\tilde H = \frac{H}{a} = \frac{\la H \ra}{a} + \tilde h.
\label{eq:scaleH}
\end{equation}
The propagator of the rescaled Higgs boson $\tilde h$ is the same as the one
of $h$
\begin{equation}
D_h(q) = D_{\tilde h}(q) = \frac{2i}{q^2 - m_H^2},
\label{eq:Dh}
\end{equation}
with
\begin{equation}
m_H^2 = 2\sigma^2 = \lambda v^2 = \frac{2\sqrt{2}\,\lambda}{G_F},
\label{eq:Mh}
\end{equation}
where we have used (\ref{eq:GF}) for the last equality;
it gets the same factor $2$ as the propagators of the goldstones
$\vec{\tilde{\mathbb P}}({\mathbb D}_1)$.

The latter are unchanged by the rescaling and given by eq.~(\ref{eq:DP});
so are the propagators of the
massive $\tilde W, \tilde Z$ gauge bosons; in the Landau gauge
\begin{equation}
D^{\mu\nu}_{W,Z}(q) = D^{\mu\nu}_{\tilde W,\tilde Z}(q)
= -i\ \frac{g^{\mu\nu} -{q^\mu q^\nu}/{q^2}}{q^2 -M_{W,Z}^2}.
\label{eq:DWZLandau}
\end{equation}
$M_W^2$ is given by eq.~(\ref{eq:MW}) and $M_Z = M_W/\cos\theta_W$, where
$\theta_W$ is the Weinberg angle.

\section{Vanishing one-loop electroweak diagrams}
\label{app:ewvanishing}

The other one-loop diagrams for $K^+ \rar \pi^+\pi^0$ and $K_s \rar
\pi^+\pi^-$ decays are all of the types drawn in fig.~10 below.

\vbox{
\figskip
\bct
\epsfig{file=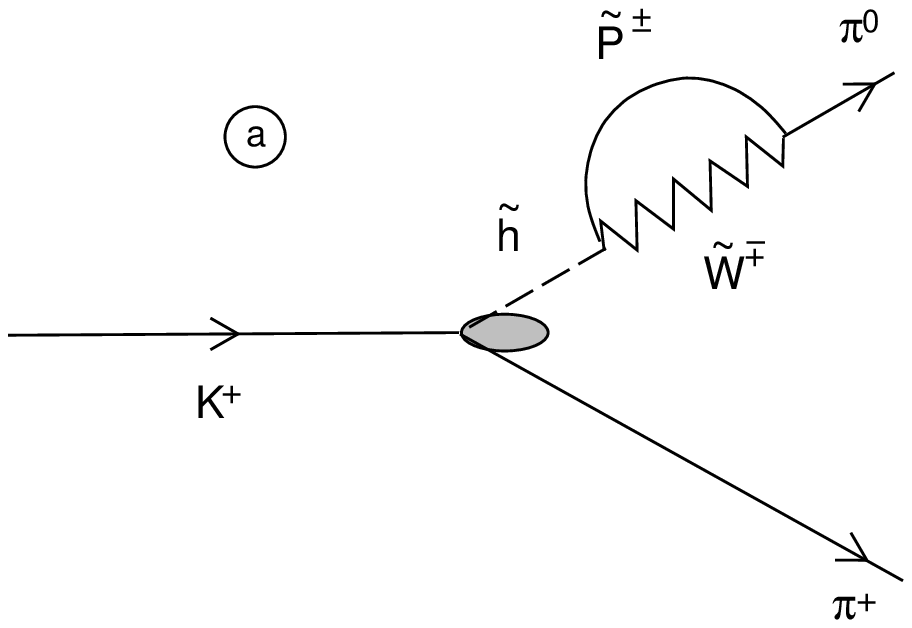,height=4truecm,width=5truecm}
\ect
\figskip
\bct
\epsfig{file=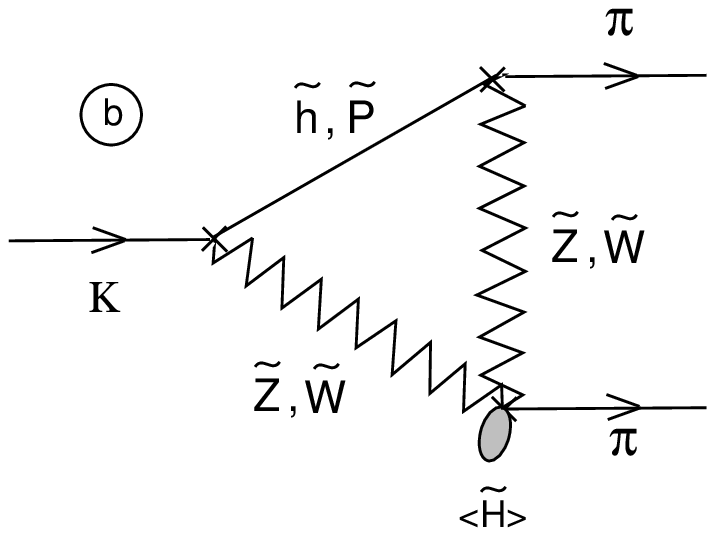,height=4truecm,width=5truecm}
\hskip 3cm
\epsfig{file=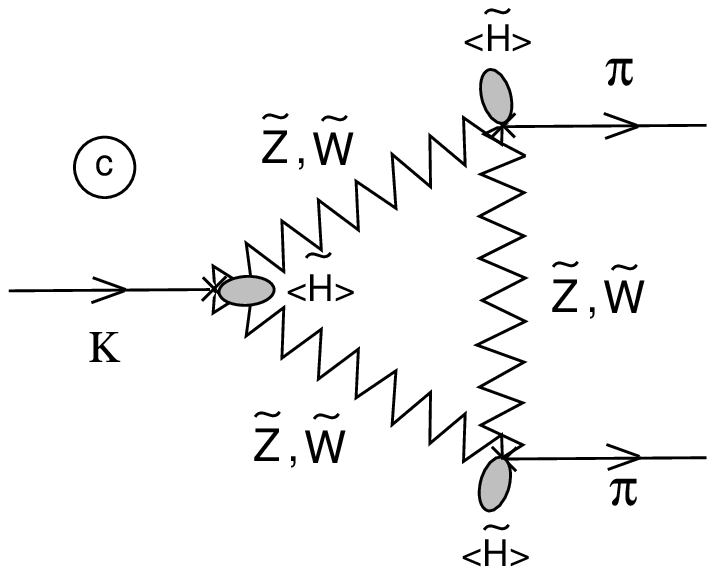,height=4truecm,width=5truecm}
\figskip
{\em Fig.~10: vanishing 1-loop electroweak diagrams for $K^+ \rar \pi^+ \pi^0$.}
\ect
}
The diagrams of figs.~10b and 10c vanish because all vertices with two gauge
bosons and one pseudoscalar meson do (all of them arise in
$(1/4)D^\mu \tilde\varphi({\mathbb D}_1) \otimes
D_\mu \tilde\varphi({\mathbb D}_1)$).

Fig.~10a vanishes as the result of the exact cancelation between
the contributions of $\ti W^+$ and $\ti W^-$.

The equivalent of fig.~10a with the $\ti W$ replaced by a $\ti Z$ does not
give a
final state with two pseudoscalars but with one scalar and one
pseudoscalar.

A similar remark eliminates other potential one-loop diagrams; so does the
fact that
the only non-vanishing vertex with three scalars involves the Higgs boson
${\mathbb S}^0({\mathbb D}_1)$.

\section{Explicit computations of one-loop electroweak diagrams}
\label{app:ewexplicit}

Neglecting $m_K^2, m_\pi^2 \ll M_{W,Z}^2, m_H^2$, one has
\begin{equation}
I_1(M^2, m_\pi^2, m_K^2) \approx
\frac{i}{16\pi^2}\frac{p.p_1}{M^2}\ \int_0^1 dy
\ \ln\frac{(1-y)m_H^2}{yM^2 +(1-y)m_H^2};
\label{eq:I1param}
\end{equation}
consequently,
\begin{equation}
I_1(M_W^2, m_\pi^2, m_K^2)-\frac{1}{c_W^2}I_1(M_Z^2, m_\pi^2, m_K^2)\approx
\frac{i}{16\pi^2}\frac{p.p_1}{M_W^2}\ \int_0^1 dy
\ \ln\frac{yM_Z^2 +(1-y)m_H^2}{yM_W^2 +(1-y)m_H^2},
\label{eq:I1W-I1Z}
\end{equation}
and, explicitly

\vbox{
\bea
&&I_1(M_W^2, m_\pi^2, m_K^2)-\frac{1}{c_W^2}I_1(M_Z^2, m_\pi^2, m_K^2)\cr
&&\approx
\frac{i}{16\pi^2}\frac{p.p_1}{M_W^2}\left(
\frac{M_Z^2}{M_Z^2 - m_H^2}\ \ln M_Z^2 -\frac{M_W^2}{M_W^2 - m_H^2}\ \ln M_W^2
+ \frac{m_H^2(M_Z^2 - M_W^2)}{(M_Z^2 - m_H^2)(M_W^2 - m_H^2)}\ \ln m_H^2\right).
\cr
&& 
\label{eq:I1W-I1Zanalyt}
\eea
}
%
%
For $I_2$, one gets
\begin{equation}
I_2(M_W^2, m_\pi^2, m_K^2)-\frac{1}{c_W^2} I_2(M_Z^2, m_\pi^2, m_K^2)\approx
\frac{i}{16\pi^2}\frac{1}{2}\frac{p.p_1}{M_W^2}\ \int_0^1 dx\ \int_0^1 dy
\ y\ln\frac{(1-y)m_H^2 + y(1-x)M_Z^2}{(1-y)m_H^2 + y(1-x)M_W^2};
\label{eq:I2W-I2Z}
\end{equation}
explicitly:
\begin{equation}
I_2(M_W^2, m_\pi^2, m_K^2)-\frac{1}{c_W^2} I_2(M_Z^2, m_\pi^2, m_K^2)
\approx
\frac{i}{16\pi^2}\frac{1}{4}\frac{p.p_1}{M_W^2}\left(
\frac{M_W^2}{M_W^2 - m_H^2}\ \ln\frac{m_H^2}{M_W^2}
-\frac{M_Z^2}{M_Z^2 - m_H^2}\ \ln\frac{m_H^2}{M_Z^2}
\right).
\label{eq:I2W-I2Zanalyt}
\end{equation}
The parametric form for the one-loop amplitude is
\bea
&&{\cal A}^{1\ loop}_{K^+\rar\pi^+\pi^0}=
-\frac{1}{4\pi^2}\ s_\theta c_\theta\ G_F^2\ f\ m_H^2\ (m_K^2 - m_\pi^2)\cr
&& \cr
&&\left(
\ \int_0^1dy
\ \ln\frac{yM_Z^2 +(1-y)m_H^2}{yM_W^2 +(1-y)m_H^2}
-\frac{1}{2}\ \int_0^1dx\ \int_0^1dy
\ y\ln\frac{(1-y)m_H^2 + y(1-x)M_Z^2}{(1-y)m_H^2 + y(1-x)M_W^2}
\right),\cr
&&
\label{eq:A1loopparam}
\eea
and its explicit analytic expression is given by eq.~(\ref{eq:A1loop}).
%
\section{Diagrams for $\mathbf{K^+ \rar \pi^+\pi^0}$ including a trilinear
$\mathbf{\Lambda}$ vertex}
\label{app:K+pi+pi0}
\subsection{Non-vanishing diagrams}
\label{subapp:nonvanK+pi+pi0}
\vbox{
\figskip
\bct
\epsfig{file=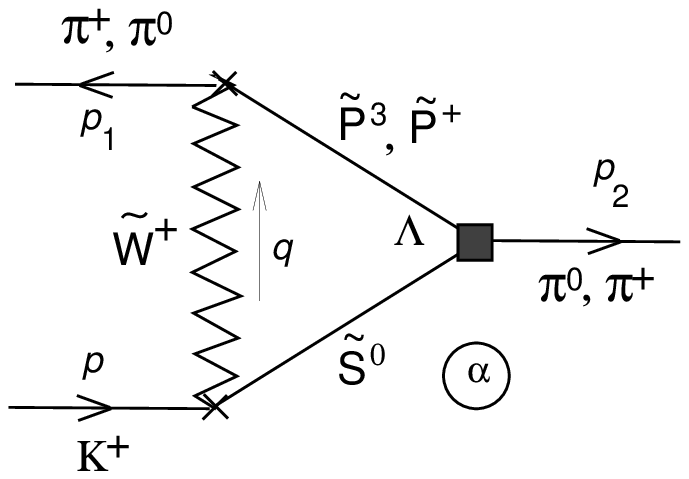,height=4truecm,width=5truecm}
\epsfig{file=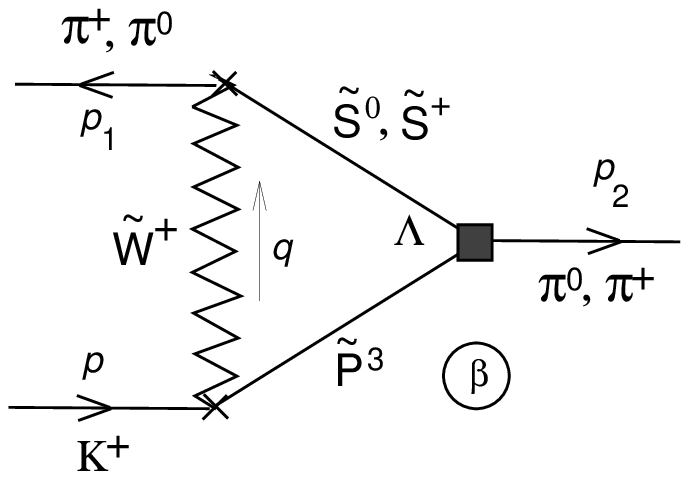,height=4truecm,width=5truecm}
\epsfig{file=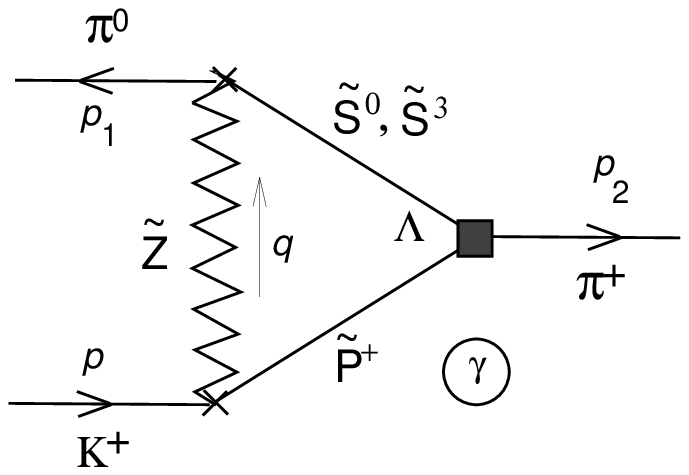,height=4truecm,width=5truecm}
\figskip
\epsfig{file=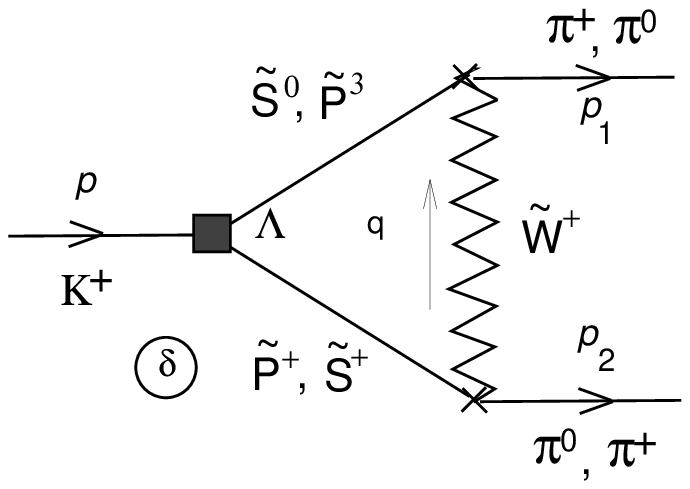,height=4truecm,width=5truecm}
\epsfig{file=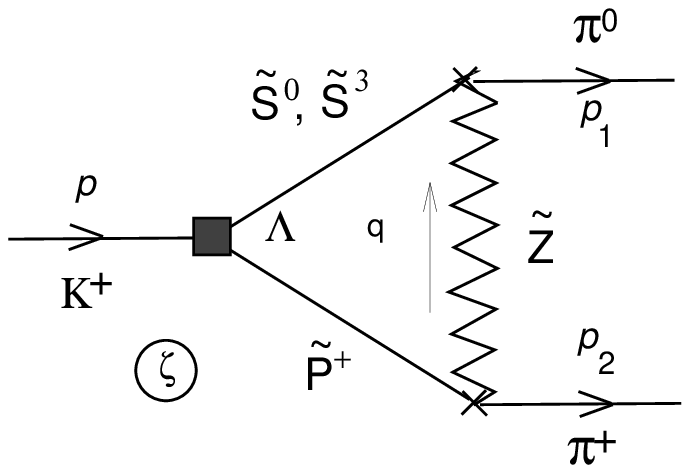,height=4truecm,width=5truecm}
\figskip
{\em Fig.~11: non-vanishing diagrams for $K^+ \rar \pi^+\pi^0$
including a trilinear $\Lambda$ coupling.}
\ect
}
%
\subsection{Vanishing or irrelevant diagrams}
\label{subapp:vanK+pi+pi0}
\vbox{
\bct
\epsfig{file=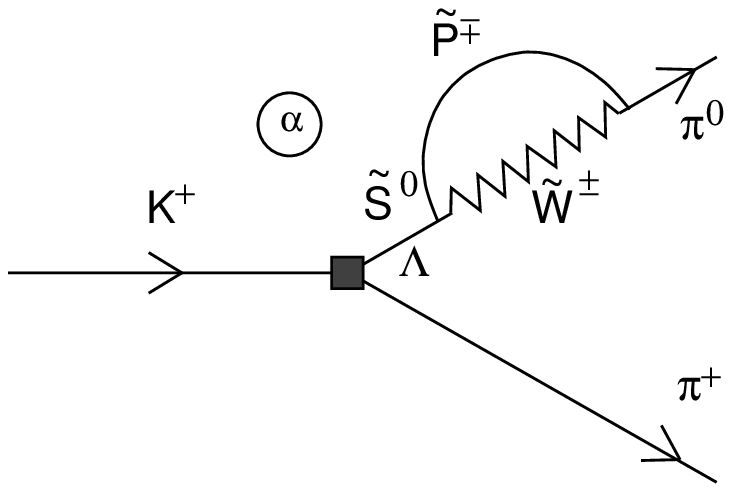,height=4truecm,width=5truecm}
\hskip 2cm
\epsfig{file=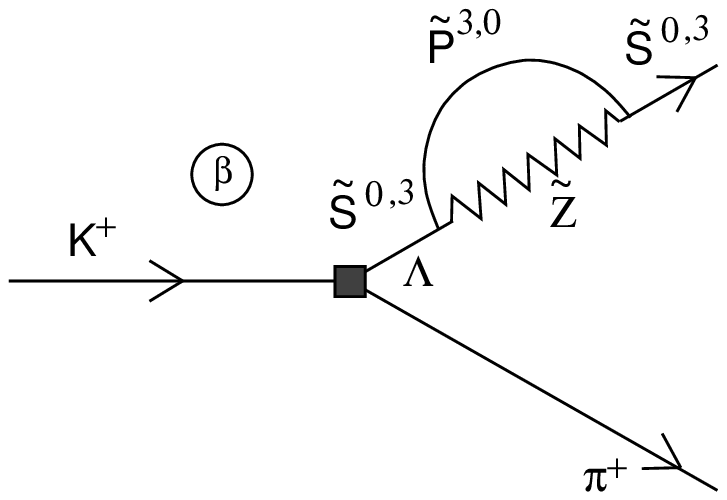,height=4truecm,width=5truecm}
\figskip
\epsfig{file=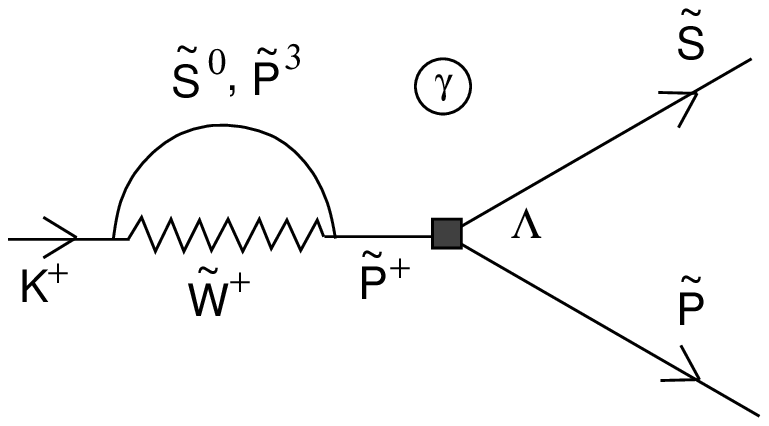,height=4truecm,width=6truecm}
\hskip 2cm
\epsfig{file=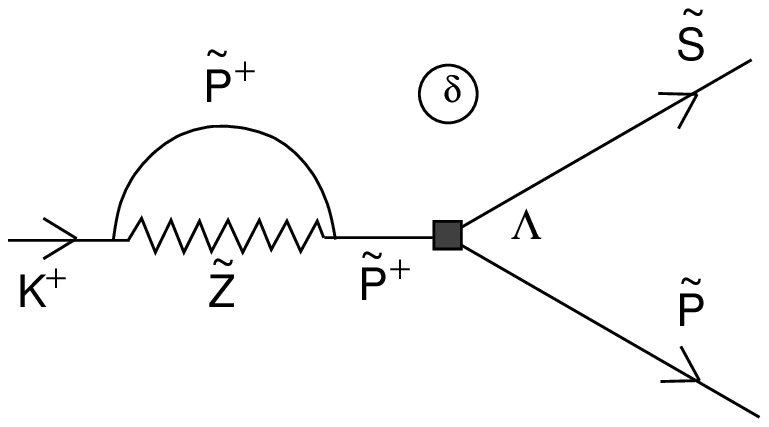,height=4truecm,width=6truecm}
\figskip
{\em Fig.~12: vanishing or irrelevant diagrams for $K^+ \rar \pi^+\pi^0$
including a trilinear $\Lambda$ coupling.}
\ect
}
The contributions of $\ti W^+$ and $\ti W^-$ in fig.~12$\alpha$ cancel.
The three other
diagrams can only yield one scalar and one pseudoscalar in the final state.
%
\section{Diagrams for $\mathbf{K_s \rar \pi\pi}$ including a trilinear 
$\mathbf{\Lambda}$ vertex}
\label{app:Kspipi}
\subsection{Non-vanishing diagrams}
\label{subapp:nonvanKspipi}
%
%
\figskip
\bct
\epsfig{file=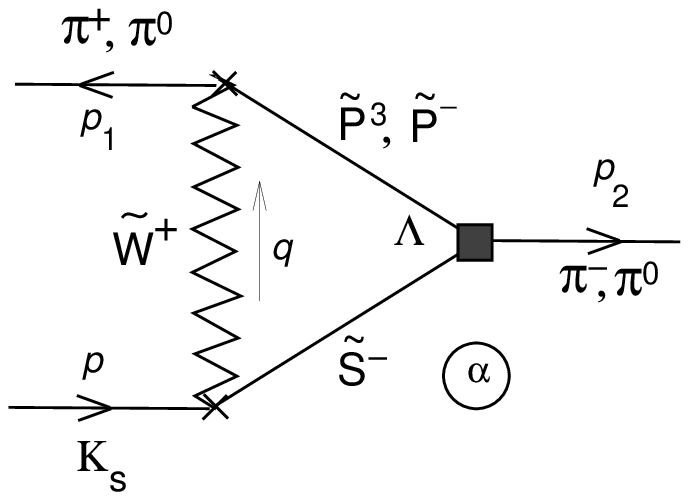,height=4truecm,width=5truecm}
\epsfig{file=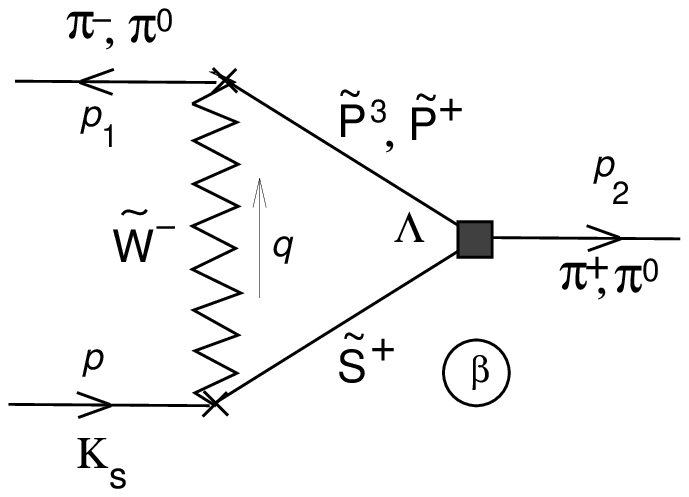,height=4truecm,width=5truecm}
\figskip
\epsfig{file=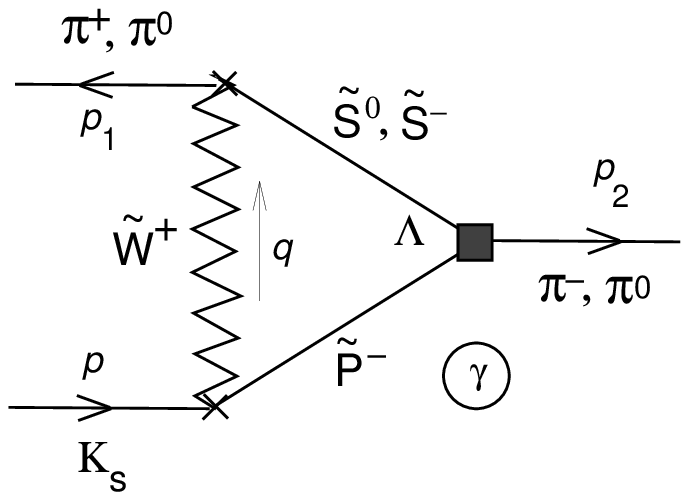,height=4truecm,width=6truecm}
\epsfig{file=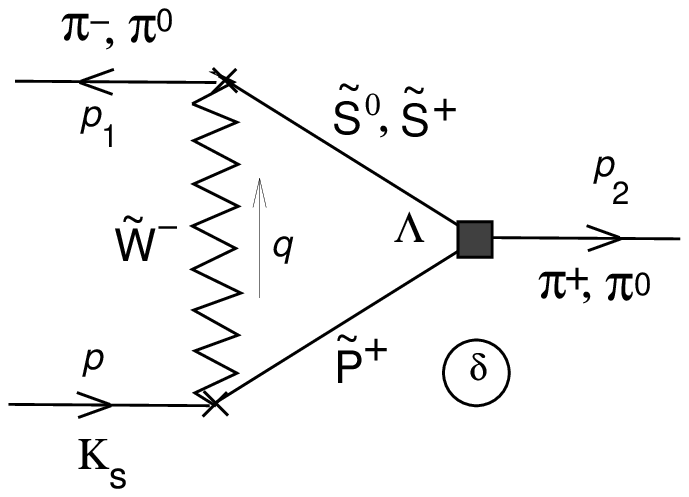,height=4truecm,width=6truecm}
\figskip
\epsfig{file=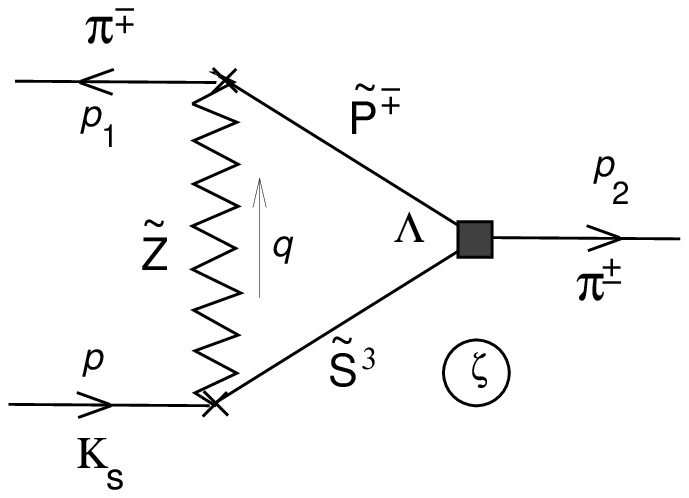,height=4truecm,width=5truecm}
\epsfig{file=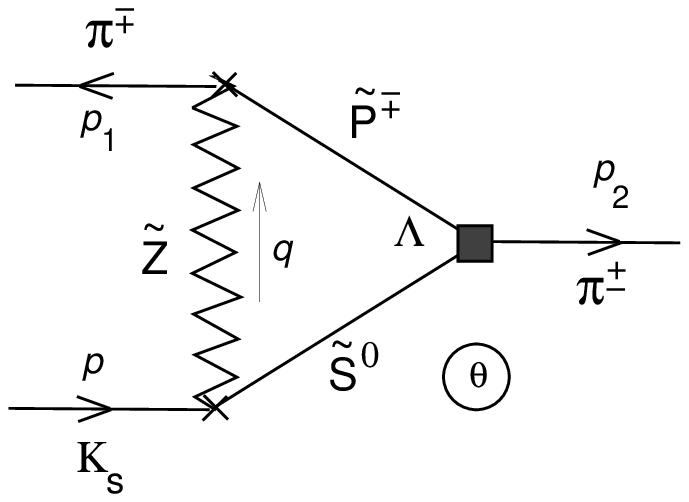,height=4truecm,width=5truecm}
\figskip
\epsfig{file=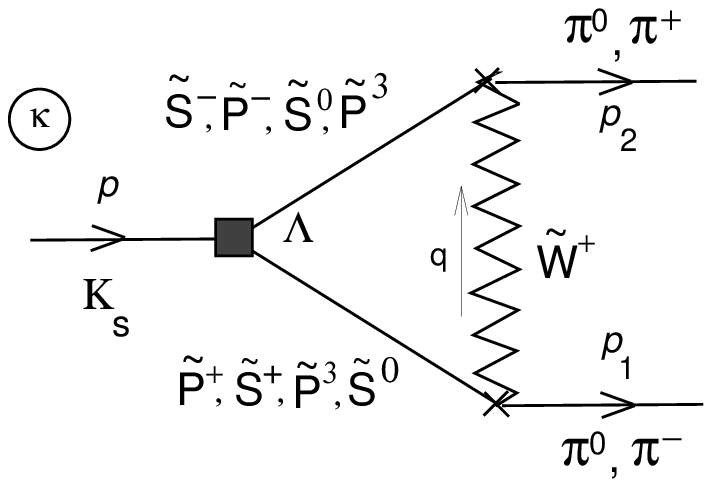,height=4truecm,width=6truecm}
\epsfig{file=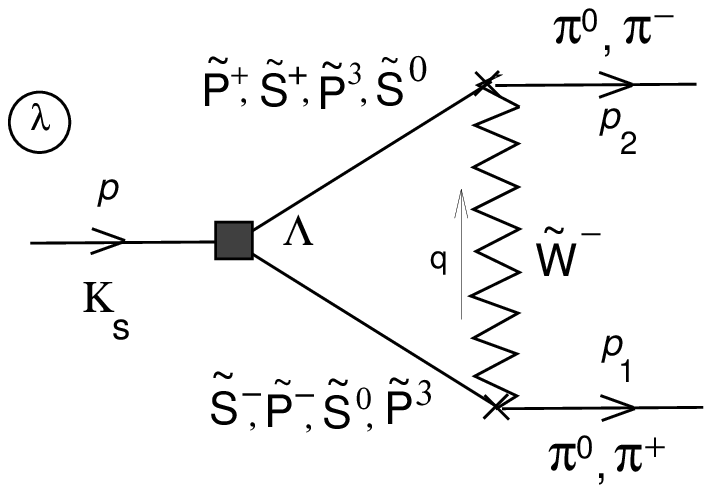,height=4truecm,width=6truecm}
\figskip
{\em Fig.~13: non-vanishing diagrams for $K_s \rar \pi\pi$
including a trilinear $\Lambda$ coupling.}
\ect
%
\subsection{Vanishing or irrelevant diagrams}
\label{subapp:vanKspipi}

They are of the same type as the ones of figs.~12; the 12$\alpha$ type of
diagram vanishes
because the contributions of $\ti W^+$ and $\ti W^-$ cancel; so do they for
the diagram of the $\gamma$ type, for which there is now canceling
contributions from $\ti W^+$ and $\ti W^-$; the $\beta$ and $\delta$ types of
diagrams are irrelevant because they can only yield one scalar and one
pseudoscalar in the final state.

\section{Linear relations between diquark operators and electroweak
eigenstates}
\label{app:relations}
\subsection{Expressing strong eigenstates in terms of electroweak
eigenstates}
\label{subapp:strongew}
\bea
(\bar u \gamma_5 u) &=& \frac{1}{2\sqrt{2}}\left(
             ({\mathbb P}^0 - i{\mathbb P}^3)({\mathbb D}_1)
               + ({\mathbb P}^0 - i{\mathbb P}^3)({\mathbb D}_2)
\right);\cr
(\bar c \gamma_5 c) &=& \frac{1}{2\sqrt{2}}\left(
    ({\mathbb P}^0 - i{\mathbb P}^3)({\mathbb D}_1)-
      ({\mathbb P}^0 - i{\mathbb P}^3)({\mathbb D}_2)
\right);\cr
(\bar d \gamma_5 d) &=& \frac{1}{2\sqrt{2}}\left(
    ({\mathbb P}^0 + i{\mathbb P}^3)({\mathbb D}_1)+
  (c_\theta^2 - s_\theta^2)({\mathbb P}^0 + i{\mathbb P}^3)({\mathbb D}_2)
  -2 s_\theta c_\theta ({\mathbb P}^0 + i{\mathbb P}^3)({\mathbb D}_3)
\right);\cr
(\bar s \gamma_5 s) &=& \frac{1}{2\sqrt{2}}\left(
    ({\mathbb P}^0 + i{\mathbb P}^3)({\mathbb D}_1)-
  (c_\theta^2 - s_\theta^2)({\mathbb P}^0 + i{\mathbb P}^3)({\mathbb D}_2)
  +2 s_\theta c_\theta ({\mathbb P}^0 + i{\mathbb P}^3)({\mathbb D}_3)
\right);\cr
(\bar u \gamma_5 d) &=& \frac{1}{2i}\left(
      c_\theta({\mathbb P}^+({\mathbb D}_1) +{\mathbb P}^+({\mathbb D}_2))
   - s_\theta ({\mathbb P}^+({\mathbb D}_3) +{\mathbb P}^+({\mathbb D}_4))
\right);\cr
(\bar d \gamma_5 u) &=& \frac{1}{2i}\left(
     c_\theta({\mathbb P}^-({\mathbb D}_1) +{\mathbb P}^-({\mathbb D}_2))
   - s_\theta ({\mathbb P}^-({\mathbb D}_3) -{\mathbb P}^-({\mathbb D}_4))
\right);\cr
(\bar u \gamma_5 s) &=& \frac{1}{2i}\left(
      c_\theta({\mathbb P}^+({\mathbb D}_3) +{\mathbb P}^+({\mathbb D}_4))
   + s_\theta ({\mathbb P}^+({\mathbb D}_1) +{\mathbb P}^+({\mathbb D}_2))
\right);\cr
(\bar s \gamma_5 u) &=& \frac{1}{2i}\left(
     c_\theta({\mathbb P}^-({\mathbb D}_3) -{\mathbb P}^-({\mathbb D}_4))
   + s_\theta ({\mathbb P}^-({\mathbb D}_1) +{\mathbb P}^-({\mathbb D}_2))
\right);\cr
(\bar u \gamma_5 c) &=& \frac{1}{2\sqrt{2}}\left(
   ({\mathbb P}^0 - i{\mathbb P}^3)({\mathbb D}_3)
   + ({\mathbb P}^0 - i{\mathbb P}^3)({\mathbb D}_4)
\right);\cr
(\bar c \gamma_5 u) &=& \frac{1}{2\sqrt{2}}\left(
   ({\mathbb P}^0 - i{\mathbb P}^3)({\mathbb D}_3)
   - ({\mathbb P}^0 - i{\mathbb P}^3)({\mathbb D}_4)
\right);\cr
(\bar d \gamma_5 s) &=& \frac{1}{2\sqrt{2}}\left(
 ({\mathbb P}^0 + i{\mathbb P}^3)({\mathbb D}_4)
+(c_\theta^2 - s_\theta^2) ({\mathbb P}^0 + i{\mathbb P}^3)({\mathbb D}_3)
+ 2s_\theta c_\theta ({\mathbb P}^0 + i{\mathbb P}^3)({\mathbb D}_2)
\right);\cr
(\bar s \gamma_5 d) &=& \frac{1}{2\sqrt{2}}\left(
 -({\mathbb P}^0 + i{\mathbb P}^3)({\mathbb D}_4)
 +(c_\theta^2 -s_\theta^2) ({\mathbb P}^0 + i{\mathbb P}^3)({\mathbb D}_3)
+ 2s_\theta c_\theta({\mathbb P}^0 + i{\mathbb P}^3)({\mathbb D}_2)
\right);\cr
(\bar d \gamma_5 c) &=& \frac{1}{2i}\left(
     c_\theta({\mathbb P}^-({\mathbb D}_3) +{\mathbb P}^-({\mathbb D}_4))
   - s_\theta ({\mathbb P}^-({\mathbb D}_1) -{\mathbb P}^-({\mathbb D}_2))
\right);\cr
(\bar c \gamma_5 d) &=& \frac{1}{2i}\left(
     c_\theta({\mathbb P}^+({\mathbb D}_3) -{\mathbb P}^+({\mathbb D}_4))
   - s_\theta ({\mathbb P}^+({\mathbb D}_1) -{\mathbb P}^+({\mathbb D}_2))
\right);\cr
(\bar c \gamma_5 s) &=& \frac{1}{2i}\left(
     c_\theta({\mathbb P}^+({\mathbb D}_1) -{\mathbb P}^+({\mathbb D}_2))
   + s_\theta ({\mathbb P}^+({\mathbb D}_3) -{\mathbb P}^+({\mathbb D}_4))
\right);\cr
(\bar s \gamma_5 c) &=& \frac{1}{2i}\left(
     c_\theta({\mathbb P}^-({\mathbb D}_1) -{\mathbb P}^-({\mathbb D}_2))
   + s_\theta ({\mathbb P}^-({\mathbb D}_3) +{\mathbb P}^-({\mathbb D}_4))
\right).\cr
&&
\label{eq:strongew}
\eea
%
\subsection{Projecting outgoing electroweak eigenstates on pion states}
\label{subapp:ewstrong}

\vbox{
\bea
&&{\mathbb P}^3({\mathbb D}_1) \ni \frac{i}{\sqrt{2}}(\bar u \gamma_5 u
   - \bar d \gamma_5 d) \equiv a\,\pi^0;\cr
&&{\mathbb P}^3({\mathbb D}_2) \ni c_\theta^2 \frac{i}{\sqrt{2}}(\bar u
            \gamma_5 u - \bar d \gamma_5 d) \equiv c_\theta^2\ a\,\pi^0;\cr
&&{\mathbb P}^3({\mathbb D}_3) \ni -s_\theta c_\theta \frac{i}{\sqrt{2}}
(\bar u \gamma_5 u - \bar d \gamma_5 d) \equiv -s_\theta c_\theta\ a\,\pi^0;
\cr
&&{\mathbb P}^3({\mathbb D}_4) \ni 0\times\pi^0;\cr
&&{\mathbb P}^0({\mathbb D}_1) \ni 0\times\pi^0;\cr
&&{\mathbb P}^0({\mathbb D}_2) \ni s_\theta^2 \frac{1}{\sqrt{2}}
(\bar u \gamma_5 u - \bar d \gamma_5 d) \equiv -is_\theta^2\ a\,\pi^0;\cr
&&{\mathbb P}^0({\mathbb D}_3) \ni s_\theta c_\theta \frac{1}{\sqrt{2}}
(\bar u \gamma_5 u - \bar d \gamma_5 d) \equiv -is_\theta c_\theta\ a\,\pi^0;
\cr
&&{\mathbb P}^0({\mathbb D}_4) \ni 0\times\pi^0;\cr
&&{\mathbb P}^+({\mathbb D}_1) \ni ic_\theta (\bar u \gamma_5 d) \equiv
           c_\theta\ a\,\pi^+;\cr
&&{\mathbb P}^+({\mathbb D}_2) \ni ic_\theta (\bar u \gamma_5 d) \equiv
           c_\theta\ a\,\pi^+;\cr
&&{\mathbb P}^+({\mathbb D}_3) \ni -is_\theta (\bar u \gamma_5 d) \equiv
           -s_\theta\ a\,\pi^+;\cr
&&{\mathbb P}^+({\mathbb D}_4) \ni -is_\theta (\bar u \gamma_5 d) \equiv
           -s_\theta\ a\,\pi^+;\cr
&&{\mathbb P}^-({\mathbb D}_1) \ni ic_\theta (\bar d \gamma_5 u) \equiv
           c_\theta\ a\,\pi^-;\cr
&&{\mathbb P}^-({\mathbb D}_2) \ni ic_\theta (\bar d \gamma_5 u) \equiv
           c_\theta\ a\,\pi^-;\cr
&&{\mathbb P}^-({\mathbb D}_3) \ni -is_\theta (\bar d \gamma_5 u) \equiv
           -s_\theta\ a\,\pi^-;\cr
&&{\mathbb P}^-({\mathbb D}_4) \ni is_\theta (\bar d \gamma_5 u) \equiv
           s_\theta\ a\,\pi^-.
\label{eq:ewstrong}
\eea
}
%

\section{$\mathbf{K^+ \rar \pi^+\pi^0}$ amplitudes}
\label{app:K+pi+pi0amps}
%

\vbox{
Fig.~11$\alpha$ yields:
\bea
{\cal S}_\alpha^{+0}(W) &=&
s_\theta c_\theta\, \Lambda \frac{ag^2}{32\sqrt{2}}
\int \frac{d^4 q}{(2\pi)^4}(2p-q)^\mu (2p_1-q)^\nu D_{\mu\nu}^W(q)\cr
&&D_{{\mathbb S}^0({\mathbb D}_4)}(p-q)
    \left(D_{{\mathbb P}^+({\mathbb D}_1)} - D_{{\mathbb P}^3({\mathbb D}_4)}
\right);
\label{eq:+0alpha}
\eea
}

\vbox{
Fig.~11$\beta$ yields:
\bea
{\cal S}_\beta^{+0}(W) &=&
s_\theta c_\theta\, \Lambda \frac{ag^2}{32\sqrt{2}}
\int \frac{d^4 q}{(2\pi)^4}(2p-q)^\mu (2p_1-q)^\nu D_{\mu\nu}^W(q)\cr
&&\left(
-\left(D_{{\mathbb S}^0({\mathbb D}_1)} + D_{{\mathbb S}^0({\mathbb D}_2)}
              \right)(p_1-q)
\left(D_{{\mathbb P}^3({\mathbb D}_1)} + D_{{\mathbb P}^3({\mathbb D}_2)}
              \right)(p-q)\right.\cr
&&\left.
     +\left(D_{{\mathbb S}^0({\mathbb D}_1)} + D_{{\mathbb S}^0({\mathbb D}_3)}
              \right)(p_1-q)
                D_{{\mathbb P}^3({\mathbb D}_3)}(p-q)\right.\cr
&&\left.
     -\left(D_{{\mathbb S}^0({\mathbb D}_4)} + D_{{\mathbb S}^+({\mathbb D}_3)}
              \right)(p_1-q)
                D_{{\mathbb P}^3({\mathbb D}_4)}(p-q)
\right);
\label{eq:+0beta}
\eea
}

\vbox{
Fig.~11$\gamma$ yields:
\bea
{\cal S}_\gamma^{+0}(Z) &=&
s_\theta c_\theta\,\frac{1}{c_W^2}\, \Lambda \frac{ag^2}{32\sqrt{2}}
\int \frac{d^4 q}{(2\pi)^4}(2p-q)^\mu (2p_1-q)^\nu D_{\mu\nu}^Z(q)\cr
&&\left(
\left(D_{{\mathbb S}^0({\mathbb D}_1)} + D_{{\mathbb S}^0({\mathbb D}_2)}
              \right)(p_1-q)
\left(D_{{\mathbb P}^+({\mathbb D}_1)} + D_{{\mathbb P}^+({\mathbb D}_2)}
              \right)(p-q)\right.\cr
&&\left.
     -\left(D_{{\mathbb S}^0({\mathbb D}_1)} + D_{{\mathbb S}^0({\mathbb D}_3)}
              \right)(p_1-q)
  \left(D_{{\mathbb P}^+({\mathbb D}_3)} - D_{{\mathbb P}^+({\mathbb
            D}_4)}\right)(p-q)
\right);
\label{eq:+0gamma}
\eea
}

\vbox{
Fig.~11$\delta$ yields:
\bea
{\cal S}_\delta^{+0}(W) &=&
-\frac{1}{2}s_\theta c_\theta\, \Lambda \frac{ag^2}{32\sqrt{2}}
\int \frac{d^4 q}{(2\pi)^4}(2p_1+q)^\mu (2p_2-q)^\nu D_{\mu\nu}^W(q)\cr
&&\left(
\left(D_{{\mathbb S}^0({\mathbb D}_1)} + D_{{\mathbb S}^0({\mathbb D}_2)}
              \right)(p_2-q)
\left(-D_{{\mathbb P}^+({\mathbb D}_3)} +D_{{\mathbb P}^+({\mathbb D}_1)}+
D_{{\mathbb P}^+({\mathbb D}_2)} \right)(p_1+q)\right.\cr
&&\left.
     -\left(D_{{\mathbb P}^3({\mathbb D}_1)} + D_{{\mathbb P}^3({\mathbb
D}_2)} \right)(p_2-q)
                D_{{\mathbb S}^+({\mathbb D}_3)}(p_1+q)\right.\cr
&&\left.
     +\left(2D_{{\mathbb S}^0({\mathbb D}_2)} - D_{{\mathbb S}^0({\mathbb
D}_3)} -D_{{\mathbb S}^0({\mathbb D}_4)} \right)(p_2-q)
    \left(D_{{\mathbb P}^+({\mathbb D}_1)}+ D_{{\mathbb P}^+({\mathbb
D}_2)}\right)(p_1+q)
\right);\cr
&&
\label{eq:+0delta}
\eea
}

\vbox{
Fig.~11$\zeta$ yields:
\bea
{\cal S}_\zeta^{+0}(Z) &=&
\frac{1}{2}s_\theta c_\theta\,\frac{1}{c_W^2}\, \Lambda \frac{ag^2}{32\sqrt{2}}
\int \frac{d^4 q}{(2\pi)^4}(2p-q)^\mu (2p_1-q)^\nu D_{\mu\nu}^Z(q)\cr
&&\left(
\left(D_{{\mathbb S}^0({\mathbb D}_1)} + D_{{\mathbb S}^0({\mathbb D}_2)}
              \right)(p_2-q)
\left(D_{{\mathbb P}^+({\mathbb D}_1)}+D_{{\mathbb P}^+({\mathbb D}_2)}
-D_{{\mathbb P}^+({\mathbb D}_3)}-D_{{\mathbb P}^+({\mathbb D}_4)}
              \right)(p_1+q)\right.\cr
&&\left.
     +\left(2D_{{\mathbb S}^0({\mathbb D}_2)} - D_{{\mathbb S}^0({\mathbb
D}_3)} - D_{{\mathbb S}^3({\mathbb D}_3)} \right)(p_2-q)
  \left(D_{{\mathbb P}^-({\mathbb D}_1)} + D_{{\mathbb P}^-({\mathbb
            D}_2)}\right)(p_1+q)
\right);
\label{eq:+0zeta}
\eea
}
%

\section{$\mathbf{K_s \rar \pi^+\pi^-}$ amplitudes}
\label{app:Kspi+pi-amps}
%
\vbox{
Fig.~13$\alpha$ yields:
\bea
{\cal S}_\alpha^{\pm}(W) &=&
s_\theta c_\theta\, \Lambda \frac{ag^2}{32\sqrt{2}}
\int \frac{d^4q}{(2\pi)^4}(2p-q)^\mu (2p_1-q)^\nu D_{\mu\nu}^W(q)\cr
&&D_{{\mathbb S}^-({\mathbb D}_4)}(p-q)
\left(D_{{\mathbb P}^3({\mathbb D}_2)}-D_{{\mathbb P}^3({\mathbb D}_3)}
   \right)(p_1-q);
\label{eq:+-alpha}
\eea
}

\vbox{
Fig.~13$\beta$ yields:
\bea
{\cal S}_\beta^{\pm}(W) &=&
s_\theta c_\theta\, \Lambda \frac{ag^2}{32\sqrt{2}}
\int \frac{d^4q}{(2\pi)^4}(2p-q)^\mu (2p_1-q)^\nu D_{\mu\nu}^W(q)\cr
&&D_{{\mathbb S}^+({\mathbb D}_4)}(p-q)
\left(D_{{\mathbb P}^3({\mathbb D}_2)}-D_{{\mathbb P}^3({\mathbb D}_3)}
   \right)(p_1-q)
\equiv {\cal S}_\alpha^{\pm}(W);
\label{eq:+-beta}
\eea
}

\vbox{
Fig.~13$\gamma$ yields:
\bea
{\cal S}_\gamma^{\pm}(W) &=&
-s_\theta c_\theta\, \Lambda \frac{ag^2}{32\sqrt{2}}
\int \frac{d^4q}{(2\pi)^4}(2p-q)^\mu (2p_1-q)^\nu D_{\mu\nu}^W(q)\cr
&&D_{{\mathbb P}^-({\mathbb D}_4)}(p-q)
\left(D_{{\mathbb S}^0({\mathbb D}_1)}+D_{{\mathbb S}^0({\mathbb D}_4)}
   \right)(p_1-q);
\label{eq:+-gamma}
\eea
}

\vbox{
Fig.~13$\delta$ yields:
\bea
{\cal S}_\delta^{\pm}(W) &=&
-s_\theta c_\theta\, \Lambda \frac{ag^2}{32\sqrt{2}}
\int \frac{d^4q}{(2\pi)^4}(2p-q)^\mu (2p_1-q)^\nu D_{\mu\nu}^W(q)\cr
&&D_{{\mathbb P}^+({\mathbb D}_4)}(p-q)
\left(D_{{\mathbb S}^0({\mathbb D}_1)}+D_{{\mathbb S}^0({\mathbb D}_4)}
   \right)(p_1-q)
\equiv {\cal S}_\gamma^{\pm}(W);
\label{eq:+-delta}
\eea
}

\vbox{
Fig.~13$\zeta$ yields:
\bea
{\cal S}_\zeta^{\pm}(Z) &=&
s_\theta c_\theta\,\frac{1}{c_W^2}\, \Lambda \frac{ag^2}{32\sqrt{2}}
\int \frac{d^4q}{(2\pi)^4}(2p-q)^\mu (2p_1-q)^\nu D_{\mu\nu}^Z(q)\cr
&&D_{{\mathbb S}^3({\mathbb D}_4)}(p-q)
\left(D_{{\mathbb P}^+({\mathbb D}_2)}+D_{{\mathbb P}^-({\mathbb D}_2)}
-D_{{\mathbb P}^+({\mathbb D}_3)}-D_{{\mathbb P}^-({\mathbb D}_3)}
   \right)(p_1-q);
\label{eq:+-zeta}
\eea
}

\vbox{
Fig.~13$\theta$ yields:
\bea
{\cal S}_\theta^{\pm}(Z) &=&
s_\theta c_\theta\,\frac{1}{c_W^2}\, \Lambda \frac{ag^2}{32\sqrt{2}}
\int \frac{d^4q}{(2\pi)^4}(2p-q)^\mu (2p_1-q)^\nu D_{\mu\nu}^Z(q)\cr
&&D_{{\mathbb S}^0({\mathbb D}_4)}(p-q)
\left(D_{{\mathbb P}^-({\mathbb D}_2)}-D_{{\mathbb P}^+({\mathbb D}_2)}
-D_{{\mathbb P}^-({\mathbb D}_3)}+D_{{\mathbb P}^+({\mathbb D}_3)}
   \right)(p_1-q);
\label{eq:+-theta}
\eea
}

\vbox{
Fig.~13$\kappa$ yields:
\begin{equation}
{\cal S}_\kappa^{\pm}(W) = 0;
\label{eq:+-kappa}
\end{equation}
}

\vbox{
Fig.~13$\lambda$ yields:
\begin{equation}
{\cal S}_\lambda^{\pm}(W) = 0.
\label{eq:+-lambda}
\end{equation}
}
%
\section{$\mathbf{K_s \rar \pi^0\pi^0}$ amplitudes}
\label{app:Kspi0pi0amps}

\vbox{
Fig.~13$\alpha$ yields:
\bea
{\cal S}_\alpha^{00}(W) &=&
s_\theta c_\theta\, \Lambda \frac{ag^2}{32\sqrt{2}}
\int \frac{d^4q}{(2\pi)^4}(2p-q)^\mu (2p_1-q)^\nu D_{\mu\nu}^W(q)\cr
&&D_{{\mathbb S}^-({\mathbb D}_4)}(p-q)
\left(D_{{\mathbb P}^-({\mathbb D}_2)}-D_{{\mathbb P}^-({\mathbb D}_3)}
   \right)(p_1-q);
\label{eq:00alpha}
\eea
}

\vbox{
Fig.~13$\beta$ yields:
\bea
{\cal S}_\beta^{00}(W) &=&
s_\theta c_\theta\, \Lambda \frac{ag^2}{32\sqrt{2}}
\int \frac{d^4q}{(2\pi)^4}(2p-q)^\mu (2p_1-q)^\nu D_{\mu\nu}^W(q)\cr
&&D_{{\mathbb S}^+({\mathbb D}_4)}(p-q)
\left(D_{{\mathbb P}^+({\mathbb D}_2)}-D_{{\mathbb P}^+({\mathbb D}_3)}
   \right)(p_1-q)
\equiv {\cal S}_\alpha^{00}(W);
\label{eq:00beta}
\eea
}

\vbox{
Fig.~13$\gamma$ yields:
\begin{equation}
{\cal S}_\gamma^{00}(W) =
-s_\theta c_\theta\, \Lambda \frac{ag^2}{32\sqrt{2}}
\int \frac{d^4q}{(2\pi)^4}(2p-q)^\mu (2p_1-q)^\nu D_{\mu\nu}^W(q)
D_{{\mathbb P}^-({\mathbb D}_4)}(p-q)
D_{{\mathbb S}^-({\mathbb D}_3)} (p_1-q);
\label{eq:00gamma}
\end{equation}
}

\vbox{
Fig.~13$\delta$ yields:
\bea
{\cal S}_\delta^{00}(W) &=&
-s_\theta c_\theta\, \Lambda \frac{ag^2}{32\sqrt{2}}
\int \frac{d^4q}{(2\pi)^4}(2p-q)^\mu (2p_1-q)^\nu D_{\mu\nu}^W(q)
D_{{\mathbb P}^+({\mathbb D}_4)}(p-q)
D_{{\mathbb S}^+({\mathbb D}_3)}(p_1-q)\cr
&\equiv& {\cal S}_\gamma^{00}(W);
\label{eq:00delta}
\eea
}

\vbox{
Fig.~13$\zeta$ yields:
\begin{equation}
{\cal S}_\zeta^{00}(Z) = 0;
\label{eq:00zeta}
\end{equation}
}

\vbox{
Fig.~13$\theta$ yields:
\begin{equation}
{\cal S}_\theta^{00}(Z) = 0;
\label{eq:00theta}
\end{equation}
}

\vbox{
Fig.~13$\kappa$ yields:
\bea
{\cal S}_\kappa^{00}(W) &=&
-\frac{1}{2}s_\theta c_\theta\, \Lambda \frac{ag^2}{32\sqrt{2}}
\int \frac{d^4q}{(2\pi)^4}(2p_1+q)^\mu (2p_2-q)^\nu D_{\mu\nu}^W(q)\cr
&&\left(
D_{{\mathbb S}^+({\mathbb D}_3)}(p_1+q)
\left(D_{{\mathbb P}^-({\mathbb D}_1)}+D_{{\mathbb P}^-({\mathbb D}_2)}
   \right)(p_2-q)\right.\cr
&&\left.+D_{{\mathbb S}^-({\mathbb D}_3)}(p_2-q)
\left(D_{{\mathbb P}^+({\mathbb D}_1)}+D_{{\mathbb P}^+({\mathbb D}_2)}
   \right)(p_1+q)
\right);
\label{eq:00kappa}
\eea
}

\vbox{
Fig.~13$\lambda$ yields:
\bea
{\cal S}_\lambda^{00}(W) &=&
-\frac{1}{2}s_\theta c_\theta\, \Lambda \frac{ag^2}{32\sqrt{2}}
\int \frac{d^4q}{(2\pi)^4}(2p_1+q)^\mu (2p_2-q)^\nu D_{\mu\nu}^W(q)\cr
&&\left(
D_{{\mathbb S}^+({\mathbb D}_3)}(p_2-q)
\left(D_{{\mathbb P}^-({\mathbb D}_1)}+D_{{\mathbb P}^-({\mathbb D}_2)}
   \right)(p_1+q)\right.\cr
&&\left.+D_{{\mathbb S}^-({\mathbb D}_3)}(p_1+q)
\left(D_{{\mathbb P}^+({\mathbb D}_1)}+D_{{\mathbb P}^+({\mathbb D}_2)}
   \right)(p_2-q)
\right)
\equiv {\cal S}_\kappa^{00}(W).
\label{eq:00lambda}
\eea
}

\newpage\null
\begin{em}

\end{em}
\end{document}